\documentclass{article}

\usepackage{arxiv}

\usepackage[utf8]{inputenc} 
\usepackage[T1]{fontenc}    
\usepackage{hyperref}       
\usepackage{url}            
\usepackage{booktabs}       
\usepackage{amsfonts}       
\usepackage{nicefrac}       
\usepackage{microtype}      
\usepackage{lipsum}		
\usepackage{graphicx}
\usepackage{doi}

\usepackage{amsmath}
\usepackage{bm}
\usepackage{bbm}
\usepackage{enumitem}
\usepackage{float}
\usepackage{caption}
\usepackage{subcaption}
\usepackage{array}
\usepackage{amsfonts}
\usepackage{arydshln}
\usepackage{booktabs}
\usepackage{url}
\usepackage[normalem]{ulem}
\usepackage{nomencl}

\title{A Probabilistic Model for Aircraft in Climb using Monotonic Functional Gaussian Process Emulators}

\author{Nick Pepper \\
	The Alan Turing Institute\\
	The British Library\\
	London, UK \\
	\texttt{npepper@turing.ac.uk} \\
	\And
	Marc Thomas \\
	NATS\\
	\And
    George {De~Ath} \\
	Department of Computer Science\\
	University of Exeter\\
	Exeter, UK \\
	\And
	Enrico Oliver \\
	Department of Computer Science\\
	University of Exeter\\
	Exeter, UK \\
	\And
	Richard Cannon\\
	NATS\\
	\And 
	Richard Everson  \\
	Department of Computer Science\\
	University of Exeter\\
	Exeter, UK \\
	\And 
    Tim Dodwell  \\
	Department of Computer Science\\
	University of Exeter\\
	Exeter, UK \\
	digiLab\\
	Exeter, UK \\
}

\begin{document}

\maketitle


\renewcommand{\shorttitle}{A Probabilistic Model for Aircraft in Climb}

\hypersetup{
pdftitle={Probabilistic Model for Aircraft in Climb},
pdfauthor={Nick Pepper, Marc Thomas, George {De~Ath}, Enrico Oliver, Richard Cannon, Richard Everson, Tim Dodwell},
pdfkeywords={Trajectory Prediction, Probabilistic Machine Learning, Functional Data Analysis, Gaussian Process Emulators, Monotonicity},
}






\begin{abstract}
Ensuring vertical separation is a key means of maintaining safe separation between aircraft in congested airspace. Aircraft trajectories are modelled in the presence of significant epistemic uncertainty, leading to discrepancies between observed trajectories and the predictions of deterministic models, hampering the task of planning to ensure safe separation. In this paper a probabilistic model is presented, for the purpose of emulating the trajectories of aircraft in climb and bounding the uncertainty of the predicted trajectory. A monotonic, functional representation exploits the spatio-temporal correlations in the radar observations. Through the use of Gaussian Process Emulators, features that parameterise the climb are mapped directly to functional outputs, providing a fast approximation, while ensuring that the resulting trajectory is monotonic. The model was applied as a probabilistic digital twin for aircraft in climb and baselined against BADA, a deterministic model widely used in industry. When applied to an unseen test dataset, the probabilistic model was found to provide a mean prediction that was 21\% more accurate, with a 34\% sharper forecast.
\end{abstract}





\maketitle

\section{Introduction}


Trajectory prediction (TP) plays an important role in the safe management of airspaces and is used by air traffic controllers to inform decision-making by predicting arrival times or detecting potential conflicts \cite{Barratt2019LearningPT}. To ensure safety, air traffic controllers strive to maintain both longitudinal and vertical separation of aircraft \cite{Paielli2009TacticalCA}. In this context vertical TP, the prediction of an aircraft's altitude with time, is especially important as controllers must devise plans that maintain separation at all times.
State-of-the-art TP methods project the future path of an aircraft through models of the flight mechanics of aircraft, in which an estimate of the present state of the aircraft is used as an initial condition for the numerical solution of a set of equations approximating the physics governing the aircraft's flight (see, e.g. \cite{bada1, bada2, bada3}). However, there is a high level of epistemic uncertainty inherent in this approach due to: the simplifications necessarily made by the model; an uncertain knowledge of the aircraft's state; lack of knowledge of the pilot's intentions \cite{bastas2020data}; and the unknown influence of environmental effects on the aircraft trajectory \cite{pang2021data}. As a consequence a mismatch between the predictions of physics-based TP methods and the actual path followed by aircraft can be observed, especially during climbs and descents \cite{climb_ga, thipphavong2013adaptive}.

\begin{figure}
\begin{center}
\includegraphics[width=0.45\textwidth]{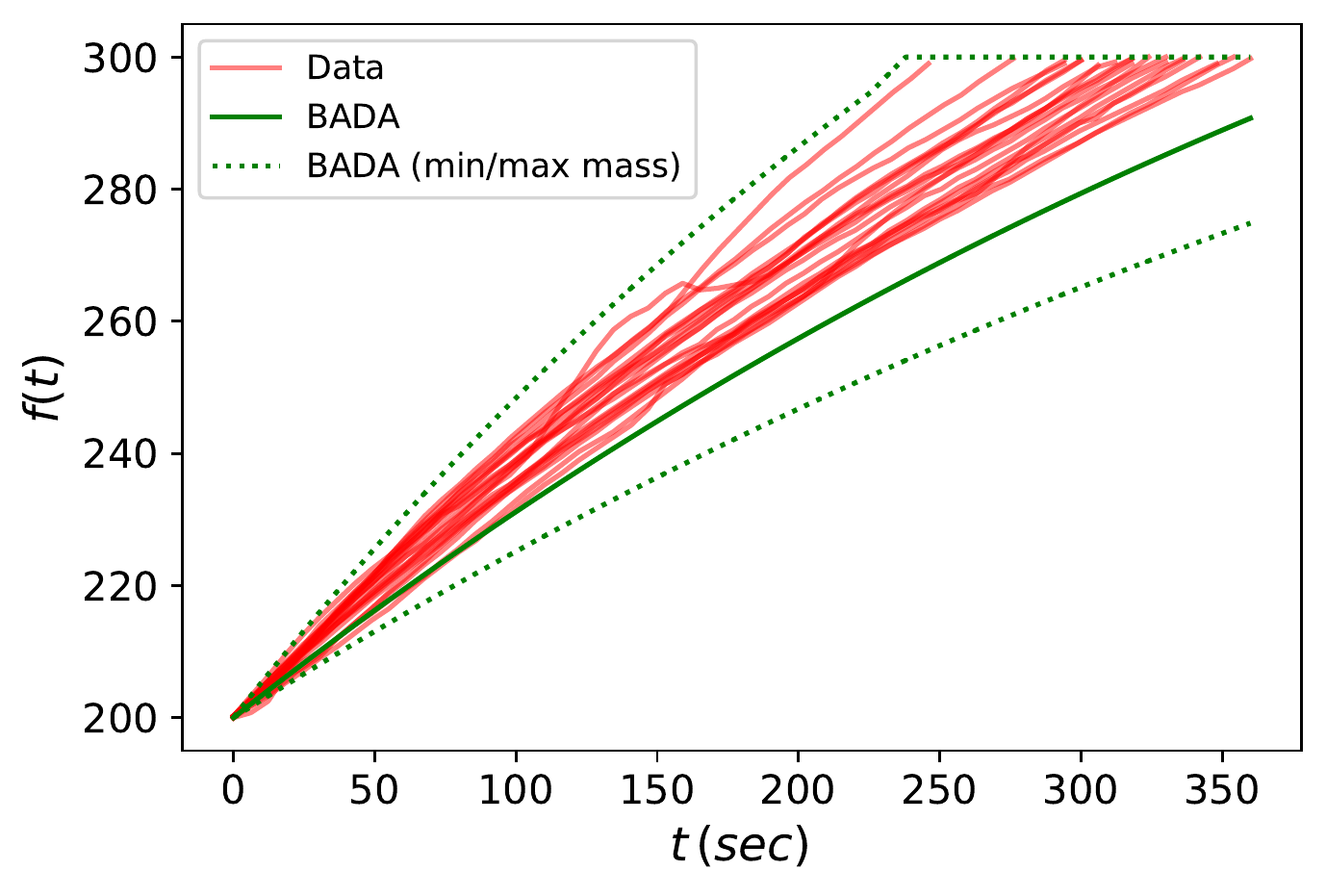}
\includegraphics[width=0.45\textwidth]{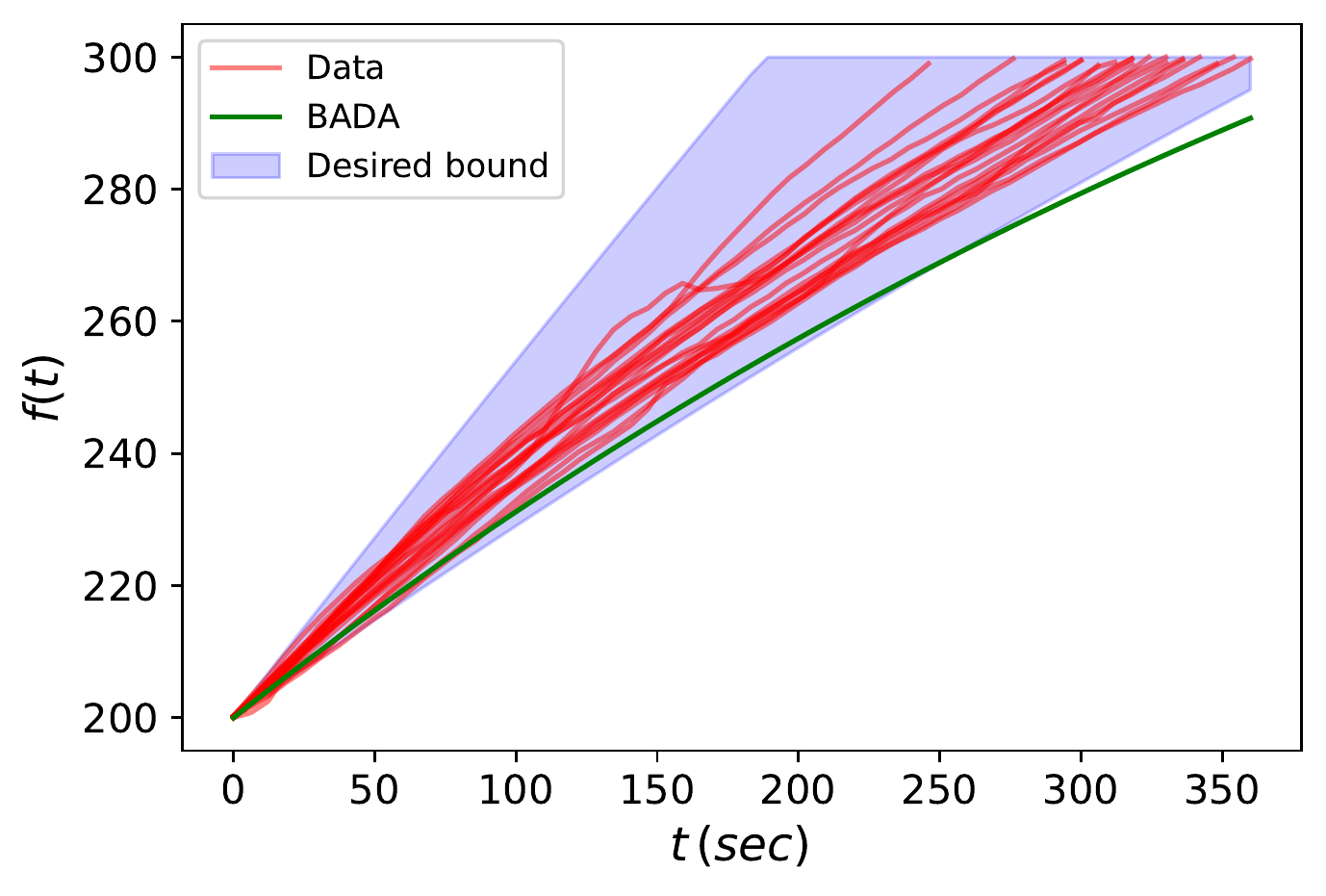}
\end{center}
\caption{An uncertainty bound for an aircraft climbing between flight levels 200 and 300, obtained from runs of a deterministic TP code with maximal and minimal mass, compared to the observed radar data (left). An illustration of the desired data-driven bounds, that better reflects the observed trajectories (right). 
}
\label{fig:cons}
\end{figure}

The observed mismatch between physics-based methods and real trajectories has motivated the application of machine learning-based methods to TP. In some respects TP is a very amenable problem for machine learning due to the large quantity of radar observations available to train models \cite{de2013machine, wang2017short, wu2022long, hernandez_ML}. In recent years, methods based on Neural Networks \cite{zhao2019aircraft, shi2018lstm} have been proposed for TP. However, there are several challenges facing machine learning methods for TP that must be addressed. Firstly, given that high levels of epistemic uncertainty contribute so significantly to model-mismatch in state-of-the-art equations-based models, a probabilistic approach would appear to be essential. Next generation TP models must be able to efficiently handle uncertainties and to clearly express the uncertainty in the model predictions \cite{pang2021data, prob_tp}. This is the motivation for approaches based on Sequential Monte Carlo sampling \cite{MC} and Gaussian Mixture Models in the TP literature \cite{Paek2020EnrouteAT,Barratt2019LearningPT}. Indeed, several recent reviews of Machine Learning methods have expressed this same point, that moving away from deterministic black-box machine learning models is necessary for the massive industrial application of machine learning \cite{ML_UQ, psaros2022uncertainty}, particularly in risk-averse industries such as aeronautics. Uncertainty estimates in TP for climbing aircraft are not generally data-driven. For instance, the left panel of Figure \ref{fig:cons} illustrates an envelope created by two runs of a deterministic TP model with minimum and maximum nominal mass, where the nominal mass refers to the aircraft mass expected by the model. However, this is an over-conservative approach. What is desired from a probabilistic TP model is illustrated on the right panel of Figure \ref{fig:cons}: a data-driven credible interval that trades off some conservatism in favour of more realistically representing the uncertainty.



In parallel with the continued refinement of TP methods, there has been a drive in recent years to explore the application of Artificial Intelligence (AI) to emulate the role of an Air Traffic Controller (ATCO) that has been motivated by the maturity of dynamic optimisation and reinforcement learning \cite{brittain2019autonomous, Ghosh_Laguna_Lim_Wynter_Poonawala_2021}. In this context, the probabilistic approach to TP appears to be particularly useful.
During training, an AI agent's plan is evaluated within a Digital Twin of an airspace. These plans must be robust to variabilities in the trajectories of aircraft, requiring a probabilistic digital twin. This offers another area of application for a probabilistic TP model \cite{Tim_NMI}.



\smallskip

The second challenge facing ML methods is to guarantee that predicted trajectories satisfy physical constraints, particularly when testing on unseen data. In the case of TP, there is a requirement that these trajectories are achievable for the performance envelope of an aircraft. Simultaneously, there is a qualitative constraint that predicted trajectories correspond to modes of aircraft operation that are observed in the real-world. In the case of aircraft in climb, a reasonable expectation is that the altitude of an aircraft will increase monotonically with time, until a target altitude is reached. Enforcing a monotonicity constraint is straightforward if the ML approach corrects the parameters of an equation-based model. For instance, Alligier et al. proposed a ML model to better predict the mass and speed intent of climbing aircraft, unknown parameters which are then fed into the Base of Aircraft Data (BADA)  physics-based model \cite{ALLIGIER201345, alligier2018learning}. However, there is a computational cost associated with the numerical solution of BADA, especially in a probabilistic approach that might require multiple model evaluations. A fast approximation to BADA is required, where model inputs are mapped directly to trajectories. However, guaranteeing that such fast approximations will themselves be monotonic is challenging. Some progress has been made in this regard in the setting of pointwise data, for example recent work with Gaussian Process Emulators (GPEs) has found correlation functions that correspond to a monotonic mean function \cite{riihimaki2010gaussian, ustyuzhaninov2020monotonic}. However, little attention has been paid in the ML literature to fast approximations with monotonic functional outputs.

\smallskip
Finally, a point prediction approach to TP, for instance by using a neural network to predict future states of an aircraft, may be limited as it does not fully exploit the spatio-temporal correlations available in the radar observations \cite{NASA_functionalTP}. Instead, a functional approach may be a more natural way to express the predicted trajectory. This is the motivating philosophy in the papers of Nicol et al. \cite{jarry2020aircraft, nicol2013functional}, in which functional Principal Component Analysis (fPCA) is used to analyse aircraft trajectories. In this work we propose a method for vertical TP using a functional Gaussian Process approach in which we marry ideas from Functional Data Analysis (FDA) with Gaussian Process Emulators. The model produces a functional estimate of a trajectory that is informed by the physics of the problem, that is, the trajectory is constrained to be monotonic. By exploiting the posterior variance, the predictive uncertainty of the method can be expressed by sampling the posterior distribution. In the following section we describe the probabilistic model, before demonstrating its application to a dataset of real flight data in Section 3, where  the model is baselined against BADA, a deterministic TP code used widely in industry.


\section{A Probabilistic model for climbing aircraft}

\noindent We propose a probabilistic model that can express the flight level, $f$, as a function of time, $t$, for an aircraft that is cleared to climb by an Air Traffic Control Officer (ATCO). Flight levels refer to the altitude at standard air pressure, expressed in hundreds of feet. A probabilistic model for the climb is denoted ${f}(t|\boldsymbol{x})$, where $\boldsymbol{x}\in \mathcal{X}\subseteq\Re^{n_x}$ represents a set of $n_x$ features that parameterise the climb. In what follows these features are considered to be time independent, consisting of the instructions issued to the aircraft by the ATCO and the available data pertaining to the aircraft's state when the command was issued. 
The model is trained using a set of radar observations containing $n_d$ monotonically increasing trajectories, $D=\{(\boldsymbol{x}^{(i)}, f^{(i)}(t))\},$ $i=1\dots n_d$. One radar sweep takes approximately 6s, while the duration of a typical climb in $D$ is on the order of 5-10 minutes. 
The method consists of two parts: firstly, trajectories in $D$ are described with a functional, monotonic representation that is parameterised by a set of hyper-parameters, $\boldsymbol{y}\in \mathcal{Y}\subseteq\Re^{n_y}$; and secondly, a set of $n_y$ Gaussian Process Emulators are defined to accomplish the probabilistic mapping $\boldsymbol{x}\rightarrow\boldsymbol{y}$. 

\subsection{Monotonic representation of functional data}
\label{section:mono}
Functional Principal Components Analysis (fPCA) is a popular tool in FDA for performing dimensionality reduction on functional data. In fPCA a function is expressed as a weighted sum of orthonormal basis functions, $\phi_i(t)$, and a mean function, $\mu(t)$:
\begin{align}
        f(t)\approx\mu(t)+\sum_{i=1}^{n_c} \boldsymbol{\alpha}_i\phi_i(t),
\end{align}
where $\boldsymbol{\alpha}\in\Re^{n_c}$ are referred to as the Principal Component scores and $n_c$ is the number of Principal Components \cite{ramsay2006functional}. fPCA has proven to provide successful representations of functional data, however, the process of finding orthogonal basis vectors is purely data-driven and is oblivious to the physical constraints on the data generating process. For instance, in the targeted application of vertical TP it is expected that $f(t)$ will be a smooth monotonic function, with constraints imposed on its derivative with respect to time, $\partial_t f$, by the performance of the aircraft. Since they are orthogonal, the fPCA modes, $\phi_i(t)$, are oscillatory. These basis functions are chosen such that they commit the smallest $L_2$ error for each $n_c$ among all possible bases.

\begin{figure}
\begin{center}
\includegraphics[width=0.49\textwidth]{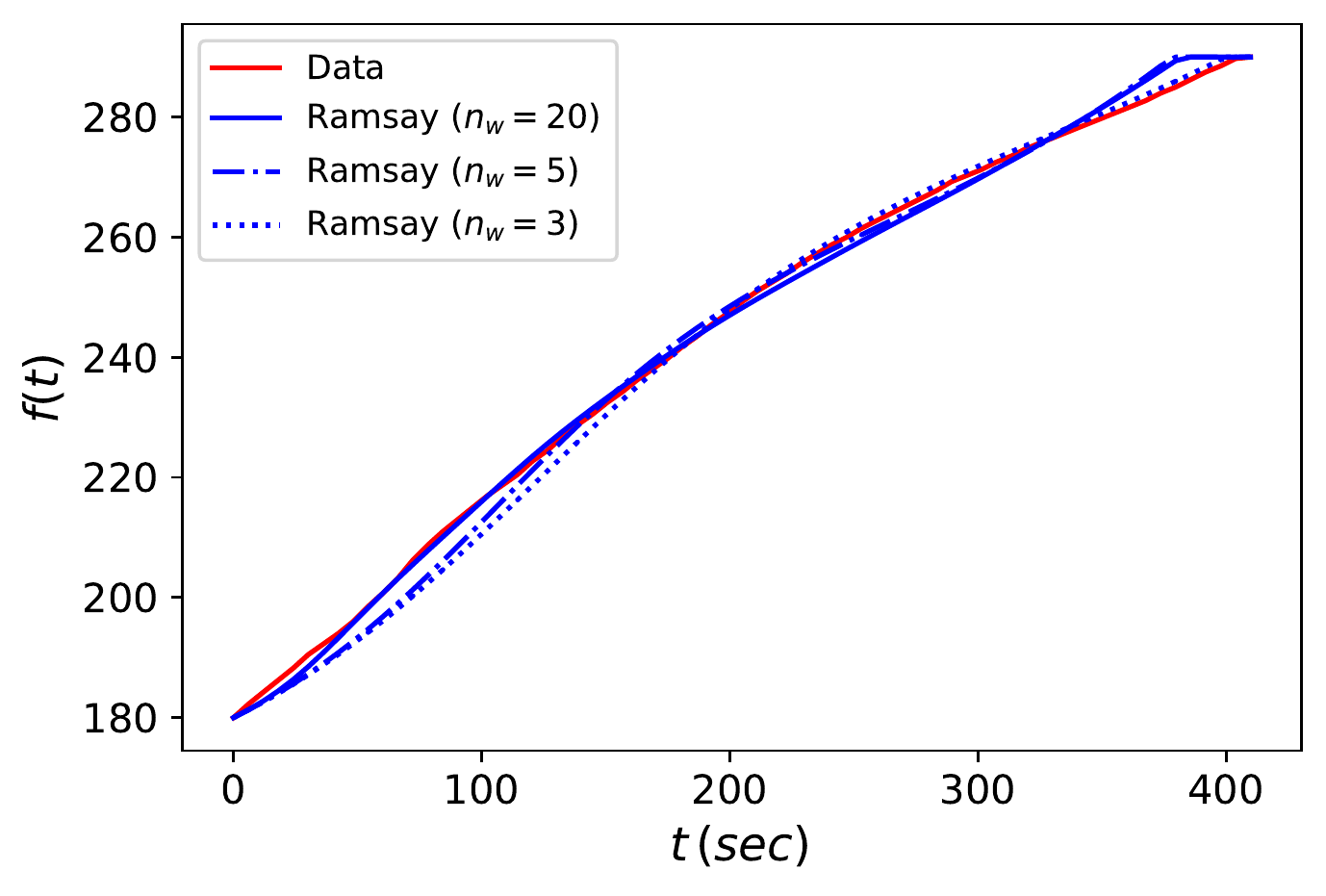}
\includegraphics[width=0.49\textwidth]{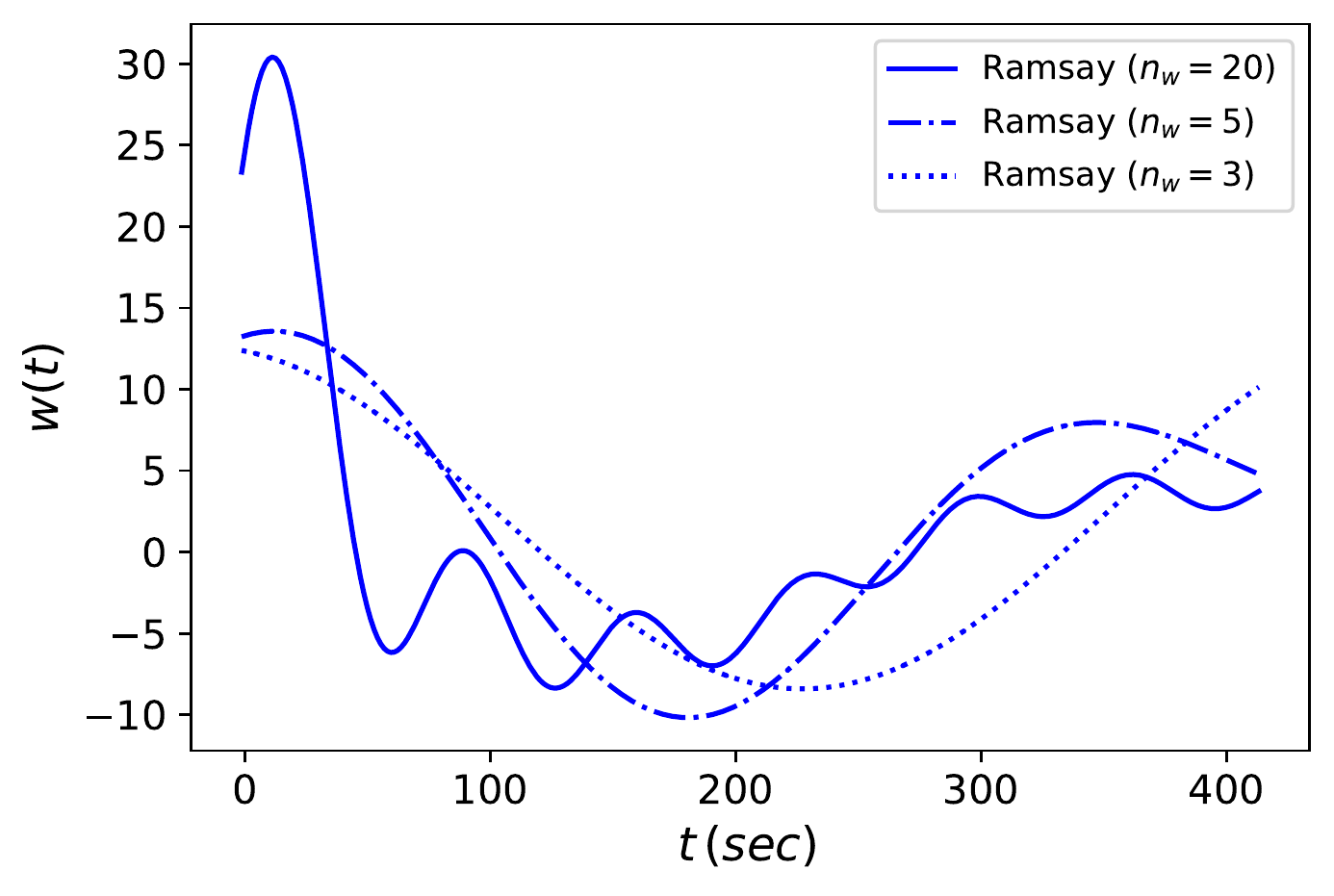}
\end{center}
\caption{Plot of a trajectory (red) and its reconstructions (blue) using the monotonic representation proposed here with varying numbers of Fourier modes (left). Associated weighting functions $w(t)$ (right).}
\label{fig:ramsay}
\end{figure}

\smallskip
In theory an infinite number of modes are required to represent monotonic functions. However, in practice the summation is truncated and it is therefore possible for the set of Principal Component scores, $\boldsymbol{\alpha}$, to correspond to a trajectory that is not monotonic even if all the trajectories in $D$ are themselves monotonic \cite{fPCAtrunc}. For this reason we, instead, employ the monotonic representation of functional data from Ramsay \cite{Ramsay2002}. Provided that $f$ satisfies the conditions that:
\begin{itemize}
    \item $\text{log}\, \partial_t f$ is differentiable;
    \item $\partial_t\,\text{log}\,\partial_t f=\partial_t^2f \slash \partial_t f$ is Lebesgue square integrable;
\end{itemize}
then such a monotonic function may be represented using an integral form:
\begin{align}
    f(t)=\beta_0+\beta_1\int_{\tau_0}^t \text{exp}\int_{\tau_0}^s w(u) \text{d}u\, \text{d}s,
    \label{mono1}
\end{align}
where $\beta_0$ and $\beta_1$ are coefficients to be determined and $w(t)$ represents a square integrable function such that $w=\partial_t^2f\slash \partial_tf$. $\tau_0$ represents the time when the manoeuvre begins. The second condition requires that the ratio of the curvature of the trajectory to its slope are bounded. Practically, satisfying these conditions requires the trajectory to be strictly monotonic (i.e. $\partial_tf>0$) and that $\partial^2_tf\rightarrow 0$ only if $\partial_tf\rightarrow0$. Given that we expect $\partial_t f$ to be continuous for an aircraft in climb, these conditions are considered reasonable for aircraft trajectories. Inspired by the work of Shin et al. \cite{Shin}, in which fPCA is used to represent this function, we cast $w(t)$ as a Fourier series:
\begin{align}
    w(t)=a_0+\sum_{i=1}^{n_w} \boldsymbol{a}_i\text{cos}(2\pi it)+\boldsymbol{b}_i\text{sin}(2\pi it),
    \label{mono2}
\end{align}
where the set of coefficients $\boldsymbol{a}, \boldsymbol{b}\in \Re^{n_w}$, $\beta_1$, and $a_0$ are determined through Stochastic Gradient Descent (SGD) for each trajectory in $D$, in which the residual sum of squares (RSS) loss between the observations and the functional representation is minimised. In what follows we enforce the initial condition $f(\tau_0=0)=f_i$, where $f_i$ represents the flight level of the first observation after the clearance to climb was issued. As a consequence $\beta_0=f_i$. Figure \ref{fig:ramsay} illustrates the reconstruction of a normalised trajectory in $D$ using this formulation and the associated $w(t)$. Note that there is some rigidity in the representation, in the example shown increasing the number of Fourier modes from 3 to 20 reduces the error in the representation of the data by 62\%, which we denote $\epsilon_{\mathcal{L}}$. However, we do not expect $\epsilon_{\mathcal{L}}\rightarrow 0$ as $n_w\rightarrow \infty$ because of the rigidity.

\smallskip
The rationale for the use of the Fourier series is to select a basis that is guaranteed to be orthogonal. This addresses a disadvantage of the method proposed by Shin, where a transformation for the B-splines used as basis functions must be found to orthogonalise them (which we note has the effect of adding oscillations into the resulting basis). A further benefit is that $w(t)$ does not need to be integrated numerically as analytical expressions are available. The Ramsay framework provides a bijective representation of the functional data in $D$. The developments in Appendix A detail the Stochastic Gradient Descent (SGD) procedure used to find the set of optimal parameters in the representation, $\hat{\boldsymbol{y}}=[\hat{\beta_1}, \hat{a}_0, \hat{\boldsymbol{a}}, \hat{\boldsymbol{b}}]^\top$,  that, when repeated for each of the trajectories in $D$, yields the set  $D_y=\{\hat{\boldsymbol{y}}^{(i)}\},\, i=1, \dots, n_d$ (with $n_y=2n_w+2$). 

\subsubsection{Dimensional Reduction with Principal Component Analysis}
Having described the $n_d$ climbs in $D$ with the monotonic framework of Ramsay, a model is desired to learn the mapping between the features $\boldsymbol{x}$ and the corresponding $2n_w+2$ parameters in this framework. To simplify this mapping, a projection of the data is performed using a Principal Component Analysis (PCA) on $D_y$.
Having performed this projection, the coefficients may be expressed using the finite sum:
\begin{align}
\boldsymbol{y}=
    \begin{bmatrix}
    \beta_1 \\
    a_0 \\
    \boldsymbol{a}\\
    \boldsymbol{b}\\
    \end{bmatrix} \approx  \sum_{k=1}^{n_c}\boldsymbol{\alpha}_k\boldsymbol{\psi}_k
    \label{eq:PCA}
\end{align}
where $\boldsymbol{\alpha}\in\Re^{n_c}$ is a vector of Principal Component (PC) scores and $\boldsymbol{\psi}_k\in\Re^{2n_w+2}\subseteq \mathcal{Y}$ is the $k$\textsuperscript{th} Principal Component. $n_c\in[1, 2n_w+2]$ is determined through an adaptive process, described in Appendix B. This projection simplifies the mapping from $\boldsymbol{x}$ because the data is now expressed in a lower-dimensional, statistically decorrelated basis (assuming $n_c<2n_w+2$). There are now two levels of approximation in the model: one in the finite sum in \eqref{mono2} and another by the finite sum in \eqref{eq:PCA}. Expressing the PC scores as a probabilistic function of $\boldsymbol{x}$, we can rewrite equations \eqref{mono1} and \eqref{mono2} to include a dependence on the features parameterising each climb:
\begin{align}
    f(t|\boldsymbol{x})=f_i+\sum_{k=1}^{n_c}\boldsymbol{\alpha}_k(\boldsymbol{x})\boldsymbol{\psi}_{1,k}\int_{\tau_0}^t \text{exp}\int_{\tau_0}^s w(u|\boldsymbol{x}) \text{d}u\, \text{d}s,
    \label{eq:mono1_alpha}
\end{align}
with
\begin{align}
 w(t|\boldsymbol{x})=\sum_{j=2}^{2n_w+2}\sum_{k=1}^{n_c}\boldsymbol{\alpha}_k(\boldsymbol{x})\boldsymbol{\psi}_{j,k}\boldsymbol{\theta}_j(t),
 \label{eq:mono2_alpha}
\end{align}
where $\boldsymbol{\psi}_{j,k}$ refers to the $j$\textsuperscript{th} element of the $k$\textsuperscript{th} Principal Component and the vector $\boldsymbol{\theta}(t)\in\Re^{2n_w+1}$ contains the Fourier modes, evaluated at time $t$, i.e.:
\begin{align}
    \boldsymbol{\theta}(t)=[1, \text{cos}(2\pi t),\dots,\text{cos}(2\pi n_wt), \text{sin}(2\pi t),\dots, \text{sin}(2 \pi n_wt)]^\top.
\end{align}
Projecting each trajectory in $D$ onto this new basis generates the set $D_\alpha=\{\boldsymbol{\alpha}^{(i)}\},\, i=1,\dots, n_d$. The sub-set $D_x=\{\boldsymbol{x}^{(i)}\},\,i=1,\dots, n_d$ and $D_\alpha$ are used as the training data for a set of independent Gaussian Process Emulators (GPEs), that are outlined in the next section.




\subsection{Gaussian Process Emulators}
Having found a monotonic functional representation of each trajectory in $D$, an interpolation function, $\Psi$, is developed for each component of $\boldsymbol{{\alpha}}$:

\begin{align}
    \hat{\boldsymbol{\alpha}}_k(\boldsymbol{{x}})={\Psi_k}(\boldsymbol{{x}})+\boldsymbol{\epsilon}_k,\; \; k=1,\dots, n_c,
\end{align}
where $\boldsymbol{\epsilon}\sim\mathcal{N}(\boldsymbol{0},\text{diag}(\boldsymbol{\sigma}_{{\epsilon}}))$ is assumed to be independent, identically distributed Gaussian noise with variance $\boldsymbol{\sigma}^2_{\epsilon}$. A set of GPEs are used to find the interpolated PCA scores ${\hat{\boldsymbol{\alpha}}}$ for a test parameter point $\boldsymbol{x}^*\subseteq \mathcal{X}$. We provide a brief summary of GPEs here, the interested reader is referred to the canonical reference books of Rasmussen and Williams \cite{GP1} and Santner et al. \cite{GP3}. A Gaussian Process prior is assumed over the regression functions, i.e.

\begin{align}
\begin{bmatrix}
\mathrm{A}_k\\
    {\hat{{\boldsymbol{\alpha}}}_k(\boldsymbol{x}^*)}
    \end{bmatrix}\sim
    \mathcal{N}\begin{pmatrix}\begin{bmatrix} \mu(X) \\ \mu(\boldsymbol{x}^*)\end{bmatrix},
    \begin{bmatrix}
    K(X,X) & K(\boldsymbol{x}^*,X)^T\\
    K(\boldsymbol{x}^*,X) & K(\boldsymbol{x}^*,\boldsymbol{x}^*)
    \end{bmatrix}\end{pmatrix},
\end{align}
where $\mu$ is a function representing the mean of the process (in what follows we set $\mu=0$). The matrix, $K$, reflects the covariances of the training data, which is organised into a $n_x\times n_d$ matrix of inputs $X=[\boldsymbol{x}^{(1)},\boldsymbol{x}^{(2)},\dots,\boldsymbol{x}^{(n_d)}]^\top$ with outputs $\mathrm{A}_k=[\boldsymbol{\alpha}_k^{(1)}, \boldsymbol{\alpha}_k^{(2)}, \dots, \boldsymbol{\alpha}_k^{(n_d)} ]$. Elements of the covariance matrix are given by:
\begin{align}
    K_{ij}(X,X)=k(\boldsymbol{x}^{(i)},\boldsymbol{x}^{(j)})+\sigma^2_{\epsilon_k}\delta_{ij},
\end{align}
where $k(.)$ represents the covariance function {and $\delta$ the Kronecker delta}. In this work squared exponential kernels were used for the covariance function. The hyperparameters for the GPE are inferred from the data, at a cost $O(n_d^3)$. To mitigate this cost when $n_d$ is large (e.g. $n_d>10^3$), we use a stochastic variational regression for the GP hyperparameters. We refer the interested reader to Blei et al. \cite{Blei_2017} and Hensman et al. \cite{hensman} for more details on this.  
A consequence of assuming a Gaussian Process prior is that the posterior predictive density is also Gaussian:

\begin{align}
    \hat{\boldsymbol{\alpha}_k}|X,\mathrm{A}_k,\boldsymbol{x}^*\sim \mathcal{N}(\hat{\mu}_k,\hat{\sigma}_k),
\end{align}
where
\begin{align}
 \hat{\mu}_k&=K(\boldsymbol{x}^*,X)^\top K(X,X)^{-1}\mathrm{A}_k,\\
 \hat{\sigma}^2_k&=K(\boldsymbol{x}^*,\boldsymbol{x}^*)-K(\boldsymbol{x}^*,X)^\top K(X,X)^{-1}K(\boldsymbol{x}^*,X).
 \nonumber
\end{align}

This is advantageous for the targeted application because the posterior distribution can be resampled for negligible cost. For a given set of features, $\boldsymbol{x}^*$, $n_{mc}$ Monte Carlo samples can be drawn, giving the set $\{\boldsymbol{\alpha}^{(i)},\dots,\boldsymbol{\alpha}^{(n_{mc}})\}$. When passed through the monotonic framework of \eqref{eq:mono1_alpha} and \eqref{eq:mono2_alpha} this yields the family of trajectories $\{f^{(i)}(t|\boldsymbol{x}^*)\},\,i=1,\dots,n_{mc}$. Having described the main features of the algorithm, in the following section we demonstrate its application to a dataset of radar observations from within a sector of UK airspace.

\section{Application to real trajectory data}
In this section we demonstrate the application of the probabilistic model to a dataset containing several thousand aircraft trajectories. This data was harvested from ATC data pertaining to a sector of UK airspace between January and February 2018. This sector contains one of the UK's busiest international airports and, as a consequence, most aircraft in the dataset are climbing to reach their cruising altitude. From the available data the most suitable features were identified by industry experts that were expected to parameterise the climbs:

\begin{enumerate}
    \item The requested change in flight level, $\Delta f$
    \item The initial flight level, $f_i$
    \item The aircraft indicated airspeed, $v_{ias}$

\end{enumerate}

\begin{figure}
\begin{center}
\includegraphics[width=0.6\textwidth]{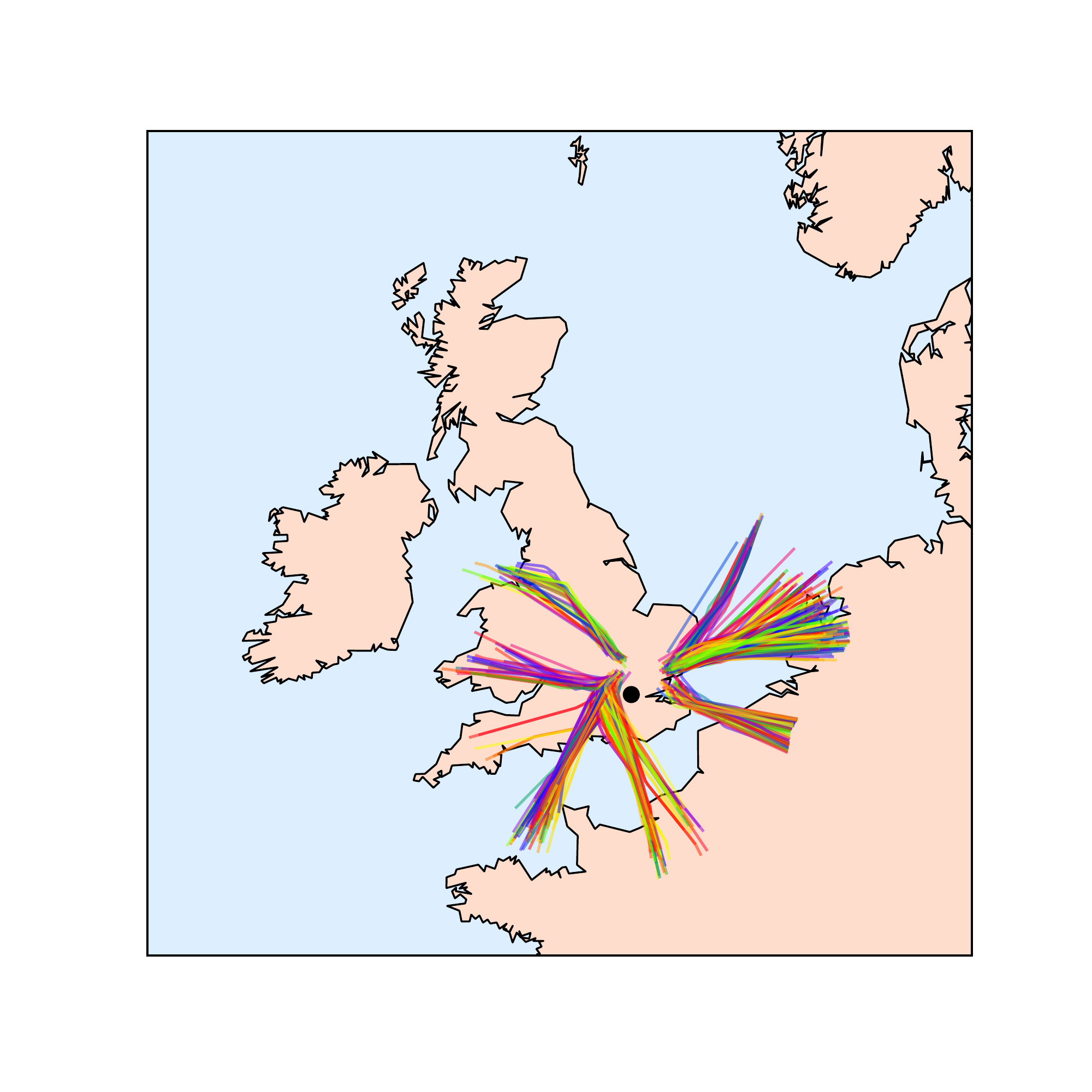}
\end{center}
\caption{Visualisation of the ground tracks of the climbs in the dataset, where each line indicates the route of a single flight in the dataset.} 
\label{fig:dataset}
\end{figure}

\noindent The three features are collected in the input vector $\boldsymbol{x}=[\Delta f, f_i, v_{ias}]^\top$. Ideally, industry experts would recommend working with the true airspeed, rather than the indicated airspeed. Such a conversion is possible provided that weather data and a model for the measurement uncertainty in $v_{ias}$ is available in the dataset. In what follows we use $v_{ias}$ but note that, should it be available, it would be possible to incorporate this additional data without altering the fundamentals of the monotonic functional method demonstrated here. Examples of parameters that are understood to significantly affect the trajectory but are not included in the dataset, because they are either unknown or proprietary, include:

\begin{itemize}
    \item The effect of variations in weather on the aircraft
    \item Mass of the aircraft
    \item True airspeed of the aircraft
    \item Aircraft performance settings
\end{itemize}
Rather than attempt to model the effects of these parameters, or assume probability  distributions for them, the effect of this epistemic uncertainty is absorbed within the posterior distribution of the GPEs. In this section we test whether such a data-driven approach can provide a credible fast approximation for the vertical trajectory of an aircraft. The data-driven uncertainty envelope is compared against one inferred from multiple runs of BADA sampled from a distribution for the aircraft mass. BADA, the deterministic TP model used as a baseline, is an energy-based model that relates to the geometrical, kinematic and kinetic aspects of the aircraft motion, allowing the aircraft performances and trajectory to be predicted. BADA is a deterministic model that is widely used by industry. Given that the output of BADA is sensitive to the mass variation of the aircraft, we contrast the probabilistic method proposed with a simple probabilistic model using BADA, in which the aircraft mass is sampled from a naive probability distribution. This simple probabilistic model is discussed further in subsection \ref{section:probBADA}. Figure \ref{fig:dataset} illustrates the ground tracks of the climbs in the dataset.

\subsection{Comparison with BADA}
The outputs of the probabilistic model were first baselined against a single, deterministic run of BADA, using the nominal mass of the aircraft. A test dataset was chosen that contained 45 trajectories, corresponding to flights from the same airline that used the same type of aircraft. The training dataset of 591 flights was used, consisting of flights with the same aircraft type and airline, but collected on 5 different days. A set of $n_{mc}=100$ posterior samples were drawn for each trajectory in the test dataset. Of the combined 636 climbs in the test and training datasets, 586 were strictly monotonic, while the remaining 50 climbs descended by no more than 1 flight level between radar observations. This discrepancy was ascribed to observational noise. 

\begin{figure}
\begin{center}
\includegraphics[width=0.49\textwidth]{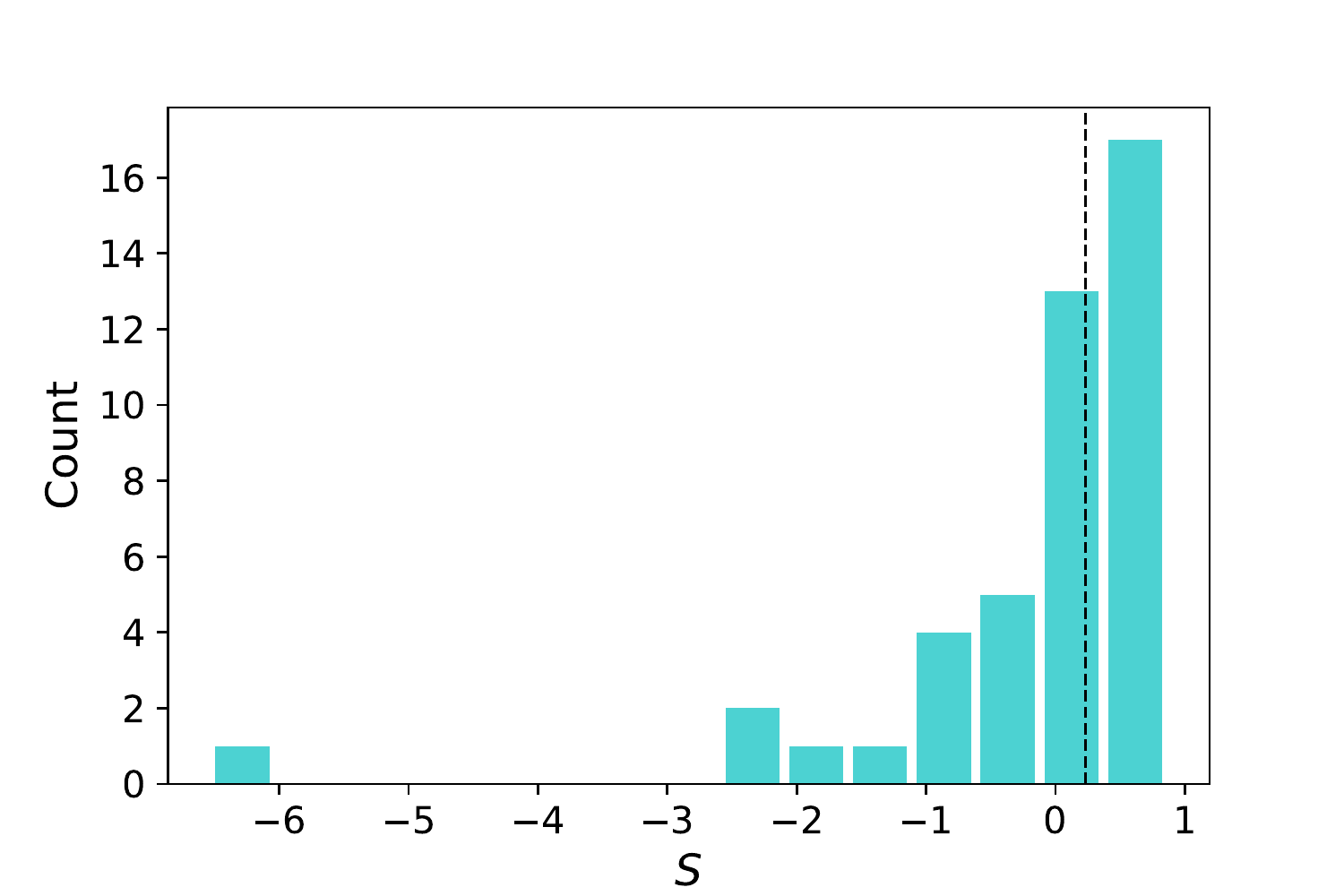}
\includegraphics[width=0.49\textwidth]{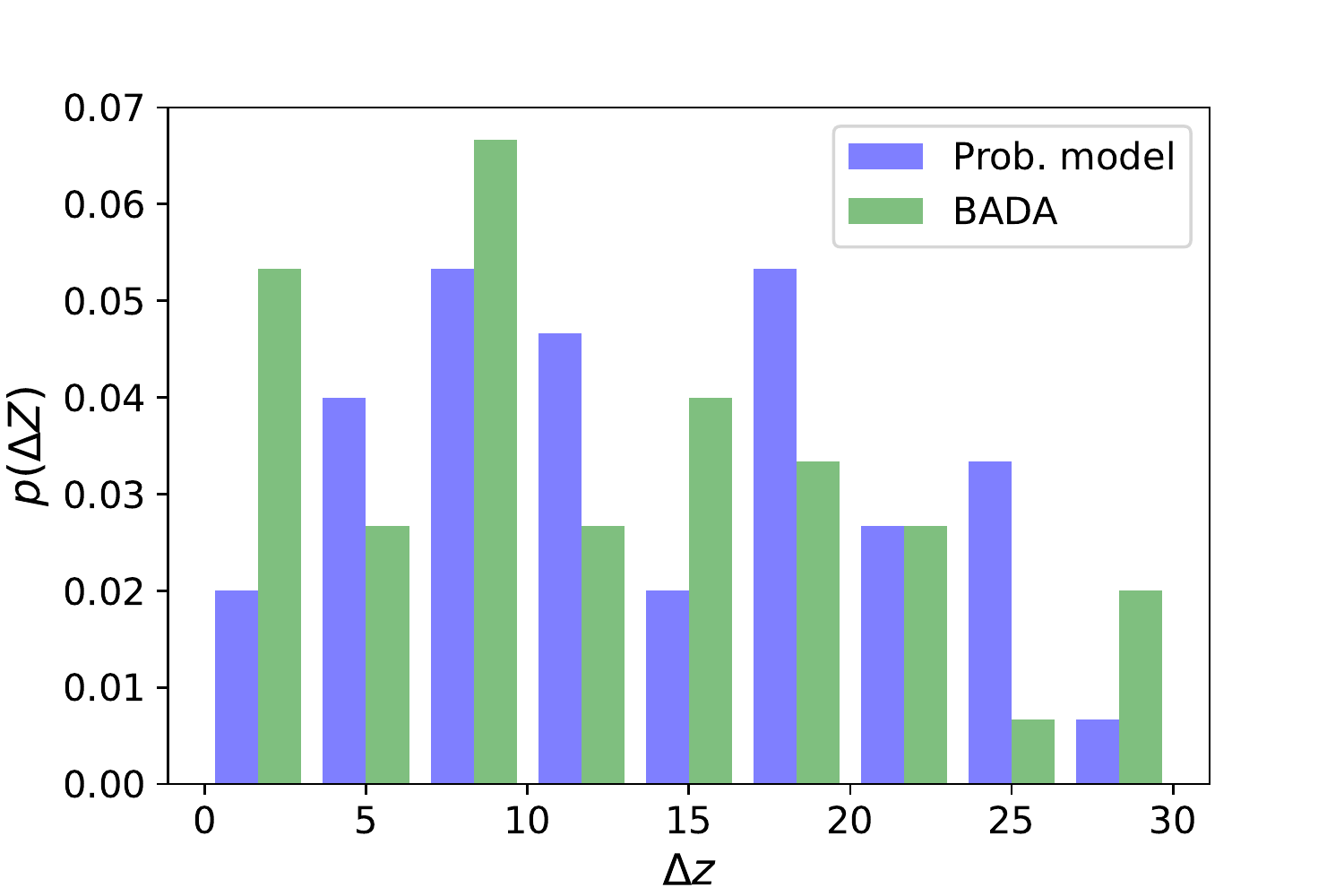}
\end{center}
\caption{Left: Histogram displaying the distribution of skill scores for
  the test dataset. The vertical line indicates the median skill. Right: Histogram displaying normalised histograms of $\Delta z$ for the probabilistic model (blue) versus BADA (green).}
\label{fig:skill_hist}
\end{figure}

A set of performance indicators were used to assess the relative skilfulness of the TP models. One of the selected indicators was the predicted time for each trajectory to achieve the target flight level, which we denote $t_a$. Taken together, a set of posterior samples provide a probabilistic forecast for $t_a$ that could be scored using the Continuous Ranked Probability Score (CRPS):
\begin{align}
    CRPS^{(i)}=\int_{-\infty}^\infty \bigg(F^{(i)}(t)- \mathbbm{1}(t-t^{(i)}_{a})\bigg)^2dt,
\end{align}
where $F^{(i)}(t)$ is the CDF for $t_a$ from the $n_{mc}$ Monte Carlo samples; $\mathbbm{1}(\cdot)$ the Heaviside step function; and $t^{(i)}_a$ the observed arrival time \cite{Gneiting}. For a deterministic forecast (i.e. a single run of BADA), the CRPS score is equivalent to the $L_2$ error. By taking the ratio of the CRPS score for the probabilistic model and the equivalent score for BADA, the relative skilfulness of the model may be expressed as a dimensionless skill score:
\begin{align}
    S^{(i)}=1-\frac{CRPS^{(i)}}{(t_a^{(i)}-t_{a,BADA}^{(i)})^2},
\end{align}
where $S^{(i)}$ represents the skill score for the $i$\textsuperscript{th} trajectory in the test set and $t_{a,BADA}^{(i)}$ the arrival time for that trajectory as predicted by BADA (with nominal mass value). Positive values of $S$ indicate that the probabilistic model is more skillful. The left panel of Figure \ref{fig:skill_hist} is a histogram displaying the skill scores for each trajectory in the test dataset, the mean skill score was found to be +32.86\%, indicating that the probabilistic model was more skilful than the deterministic run of BADA for this performance indicator.

The offset in the predictions at the arrival time, $\Delta z$, was also used as a performance indicator. From an ATC perspective this is a useful quantity because it is connected to the flight level occupancy of the aircraft. For the $i$\textsuperscript{th} trajectory in the test dataset, with observed arrival time, $t_a$ and estimated arrival time $\hat{t}_a$, this quantity is denoted:
%
%
%
\begin{align}
    \Delta z^{(i)}=
    \begin{cases}
    (f_i+\Delta f)-f({t}_a|\boldsymbol{x}^{(i)})&\; \textrm{if} \;t_a\leq \hat{t}_{a},\\
    (f_i+\Delta f)-f^{(i)}(\hat{t}_a)&\; \textrm{if} \;t_a> \hat{t}_{a}
    \end{cases}.
\end{align}
Lower values of $\Delta z$ indicate that a model has more accurately predicted the flight level occupancy at the arrival time. The right panel of Figure \ref{fig:skill_hist} displays two (normalised) histograms for $\Delta z$, with $\Delta z$ calculated for each trajectory in the test dataset for BADA (green) and for the mean of the probabilistic model (blue). As can be seen, these discrepancies are comparable, and the mean discrepancy across all 45 test flights is tabulated in Table \ref{Table:1}. This Table also tabulates the Mean Absolute Error (MAE) between the means of the models and the observed trajectory. The mean prediction of the probabilistic model offered a 21\% improvement in accuracy for this test dataset.  

\begin{figure}
\begin{center}
\includegraphics[width=0.49\textwidth]{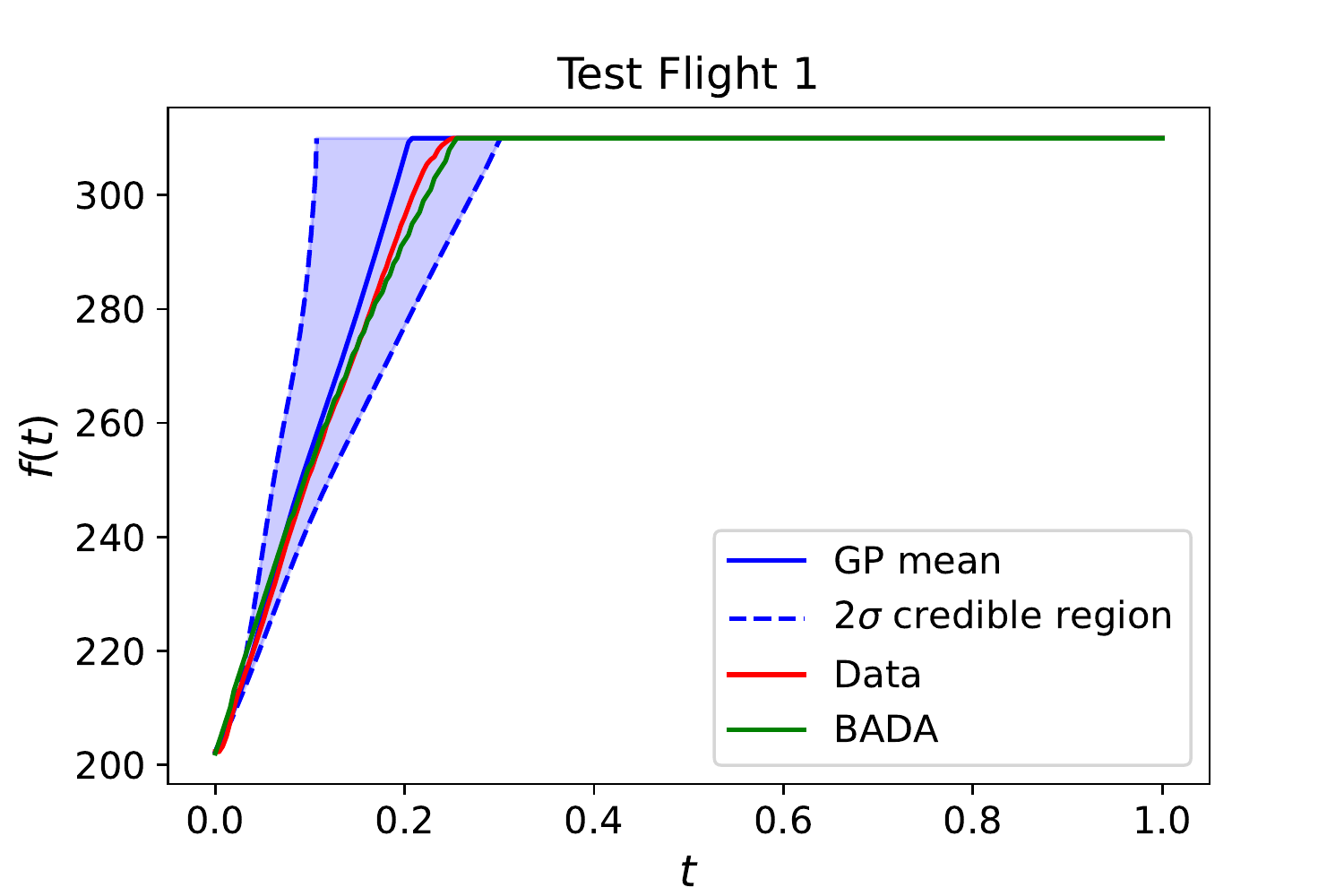}
\includegraphics[width=0.49\textwidth]{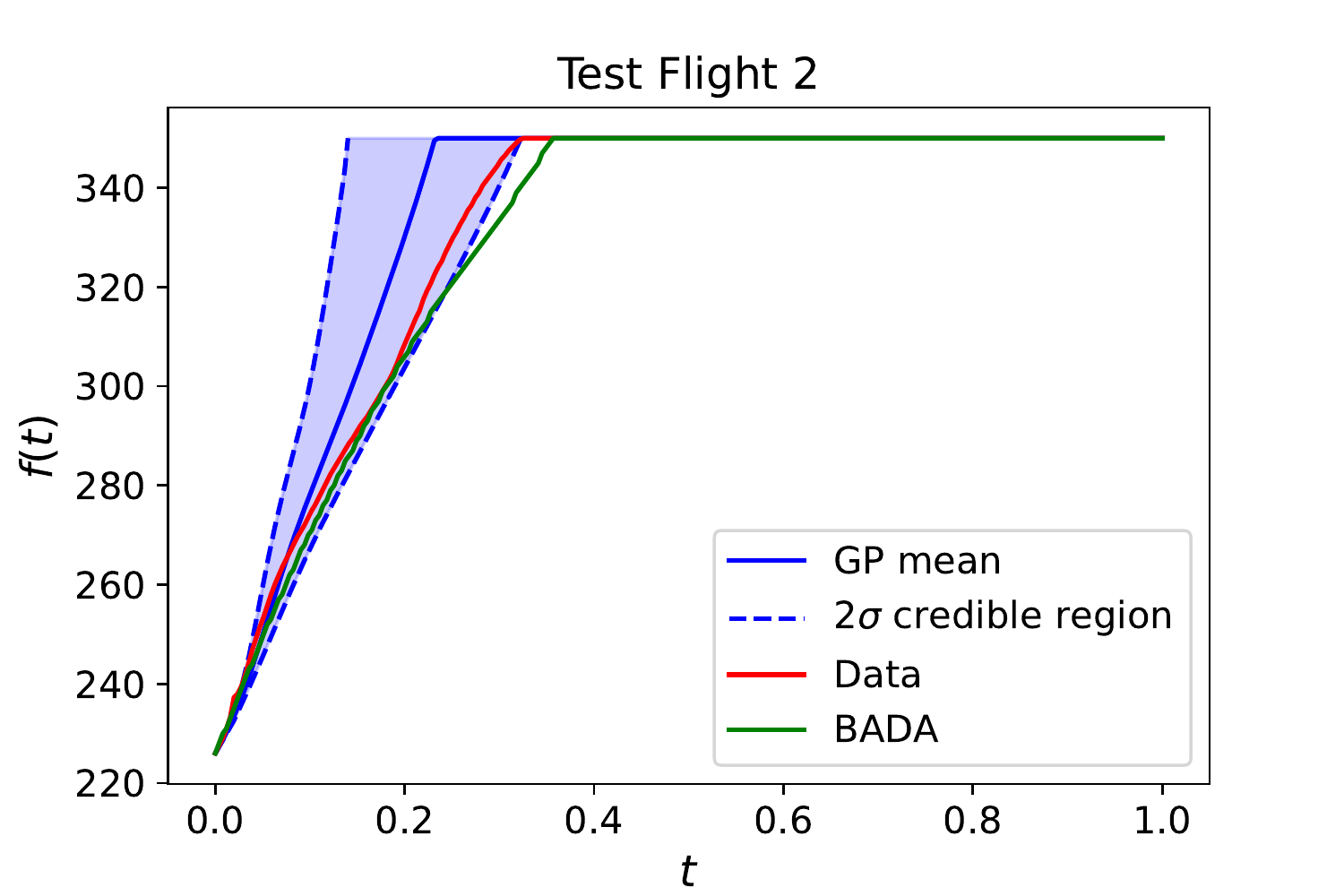}
\includegraphics[width=0.49\textwidth]{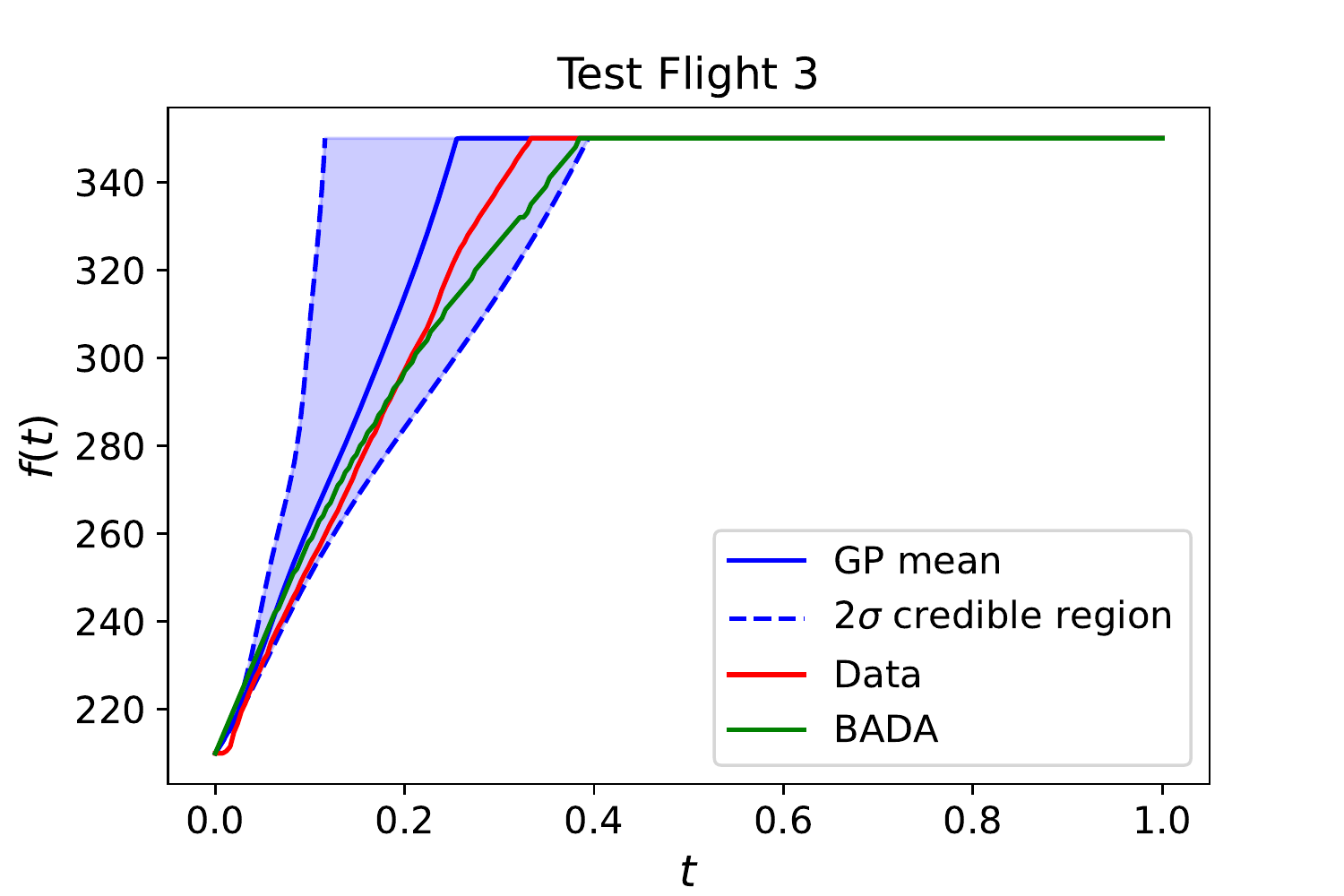}
\includegraphics[width=0.49\textwidth]{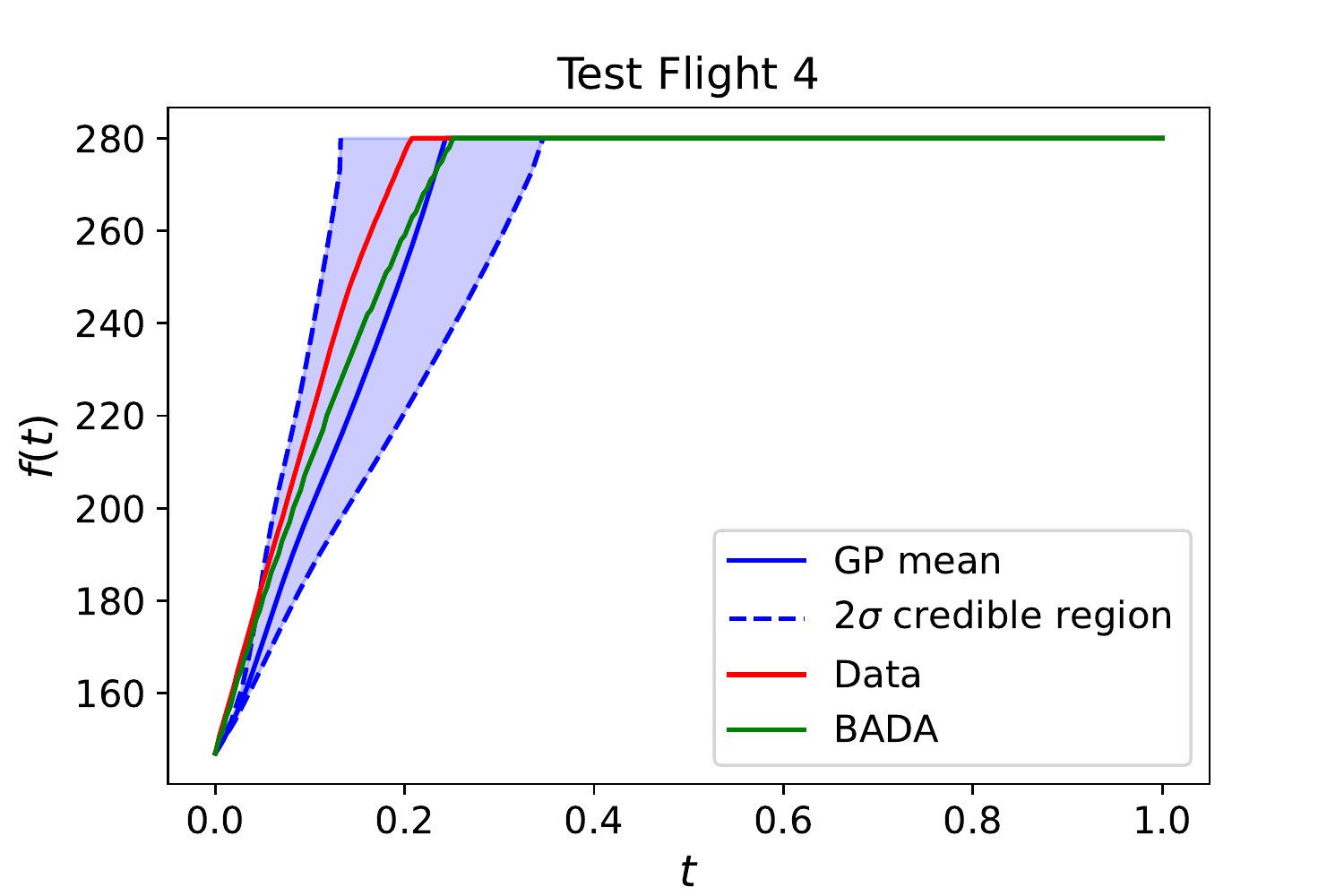}
\includegraphics[width=0.49\textwidth]{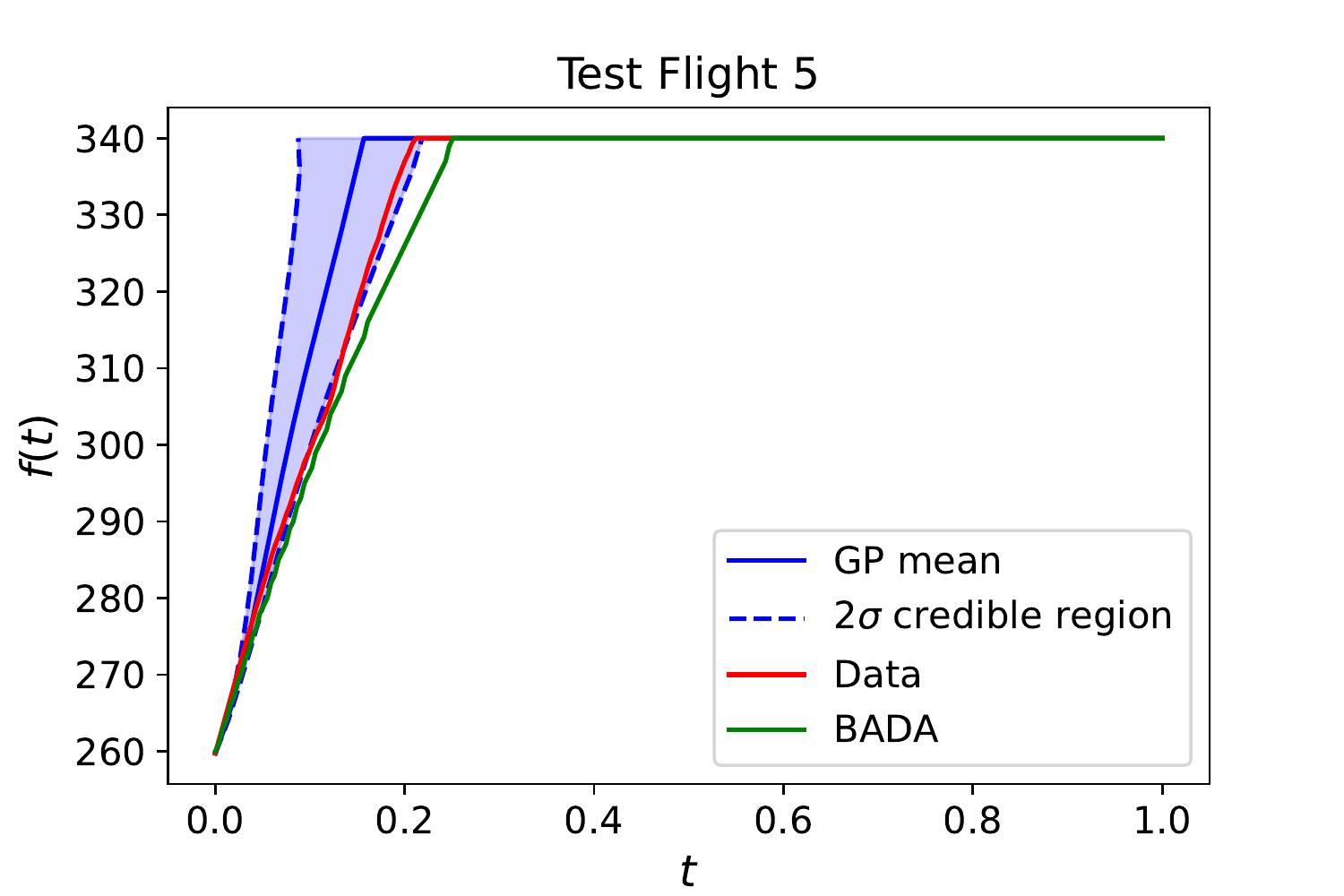}
\includegraphics[width=0.49\textwidth]{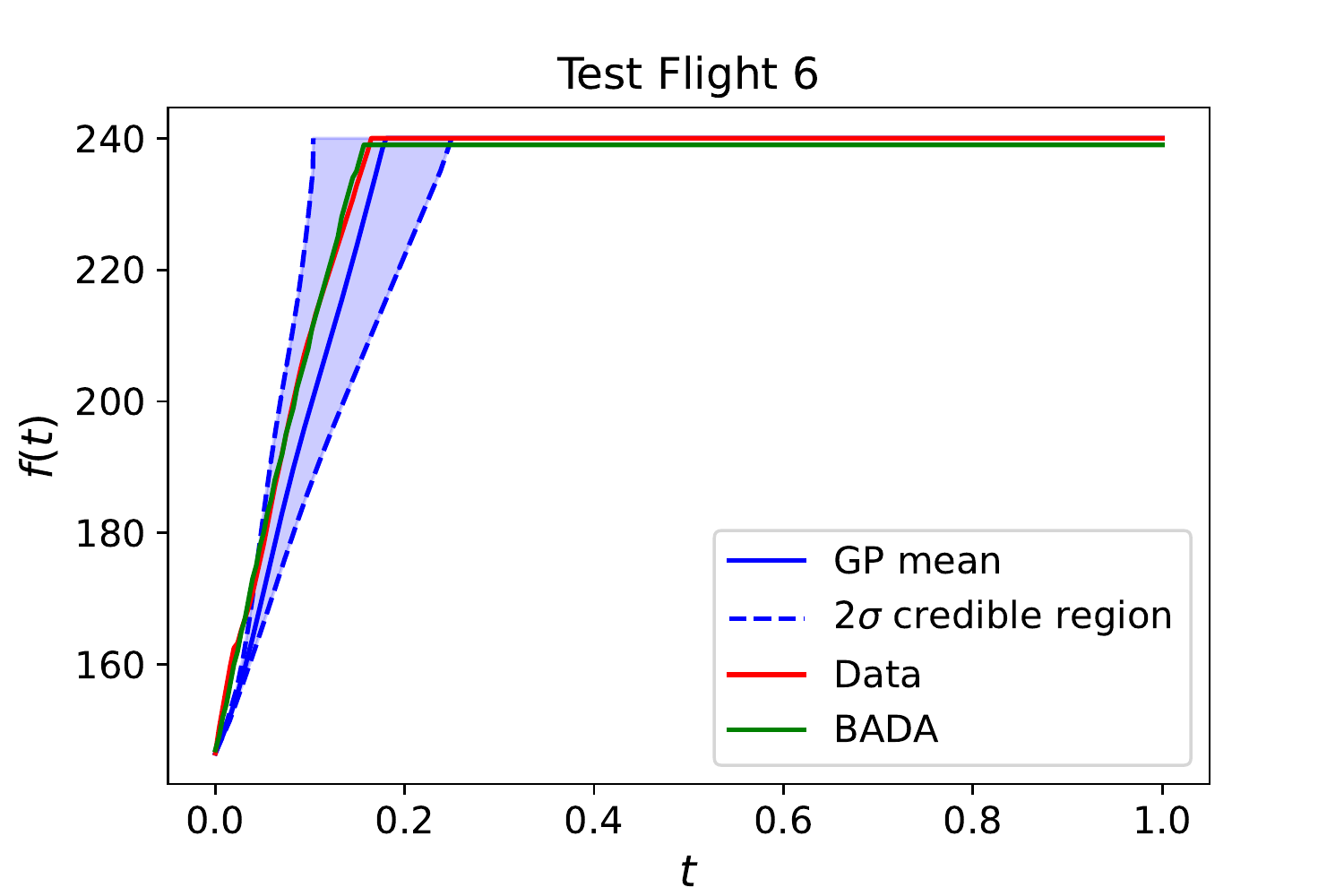}
\includegraphics[width=0.49\textwidth]{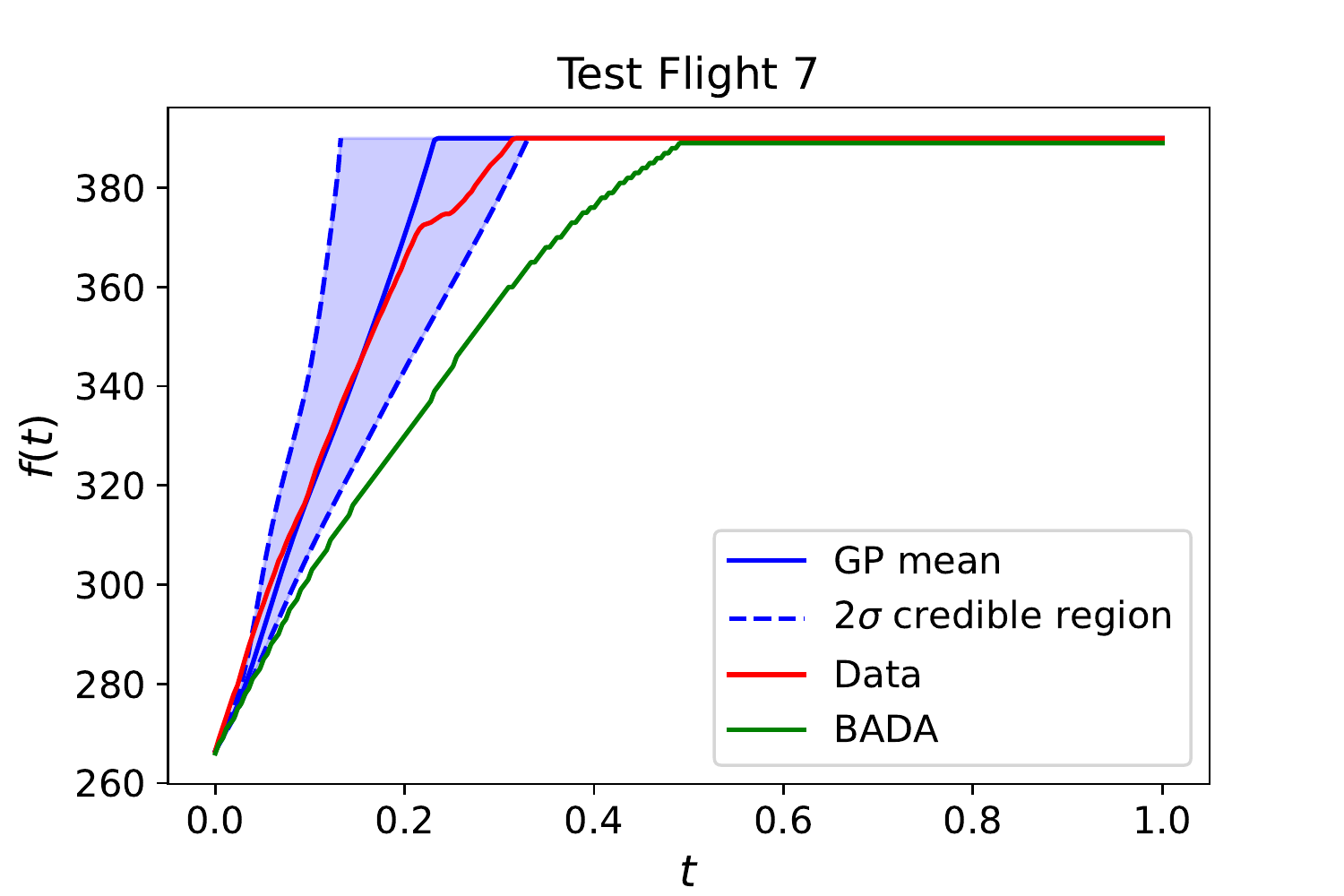}
\includegraphics[width=0.49\textwidth]{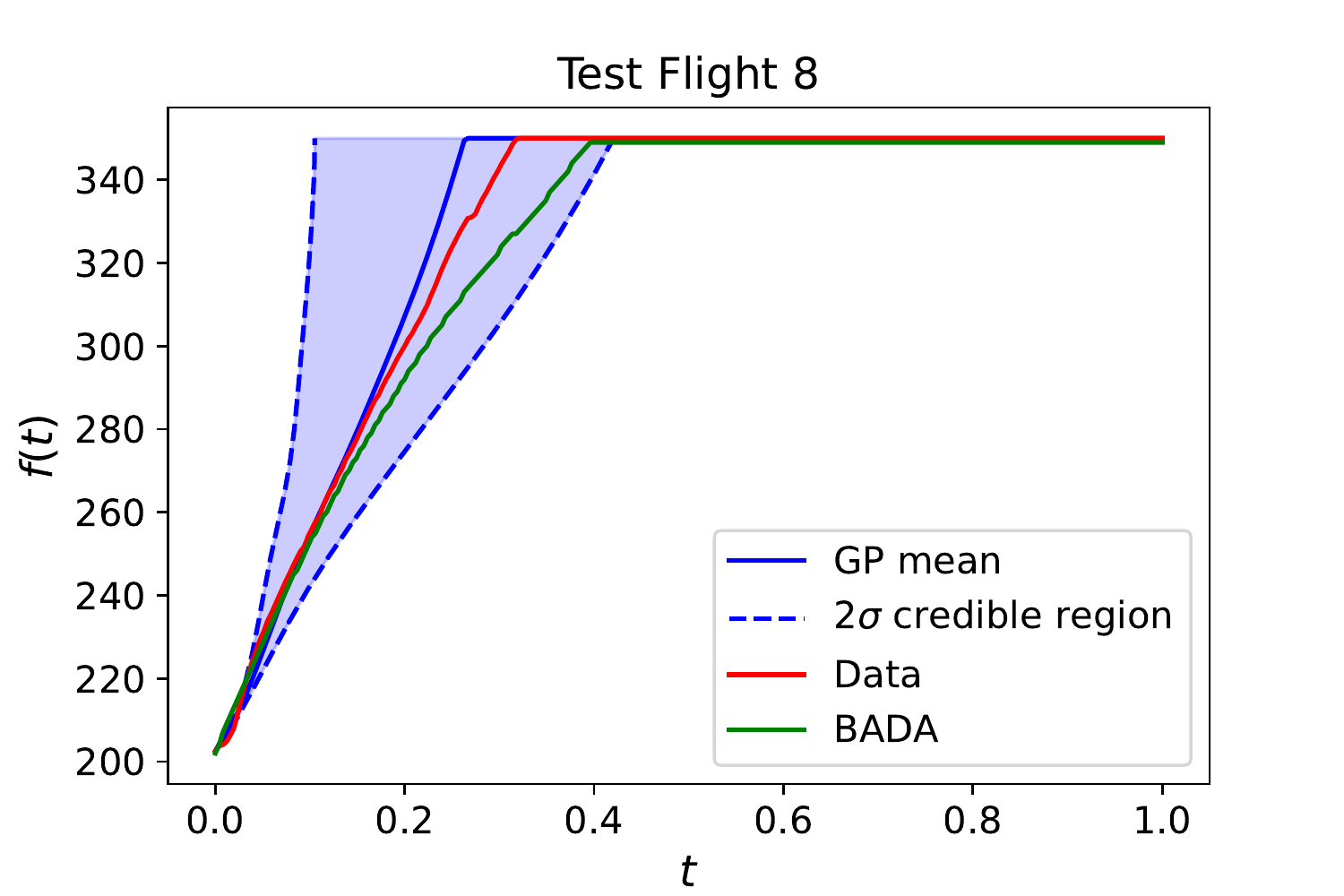}
\end{center}
\caption{Radar data for an unseen trajectory (red), compared to the predicted trajectory by BADA (green), and the mean of the probabilistic digital twin (blue) for eight trajectories in the test dataset. The dashed lines indicate the 2$\sigma$ credible region from 100 posterior samples. Note that the credible regions are drawn with respect to arrival time, rather than flight level.}
\label{fig:test_points}
\end{figure}

Figure \ref{fig:test_points} visualises the output of the probabilistic model for eight of the trajectories in the test set, displaying a range of skill scores. 
The red line indicates the observed radar data, the green line the solution of BADA for the requested climb, and the blue line the mean prediction of the generative model. The variance in the arrival time at intermediate flight levels is used to construct a 2$\sigma$ credible region for the probabilistic model. Note that in some instances, such as the seventh test flight, the data-driven credible interval is sufficiently tight that it contains the observed trajectory but not the deterministic run of BADA. The skill scores of each test flight are tabulated in Table \ref{Table:2}. In the next section, the probabilistic model is compared against a probabilistic implementation of BADA and the two metrics used to assess the credible intervals are introduced.  

\subsection{Comparison with BADA using assumed aircraft mass distribution}
\label{section:probBADA}
\begin{figure}
\begin{center}
\includegraphics[width=0.49\textwidth]{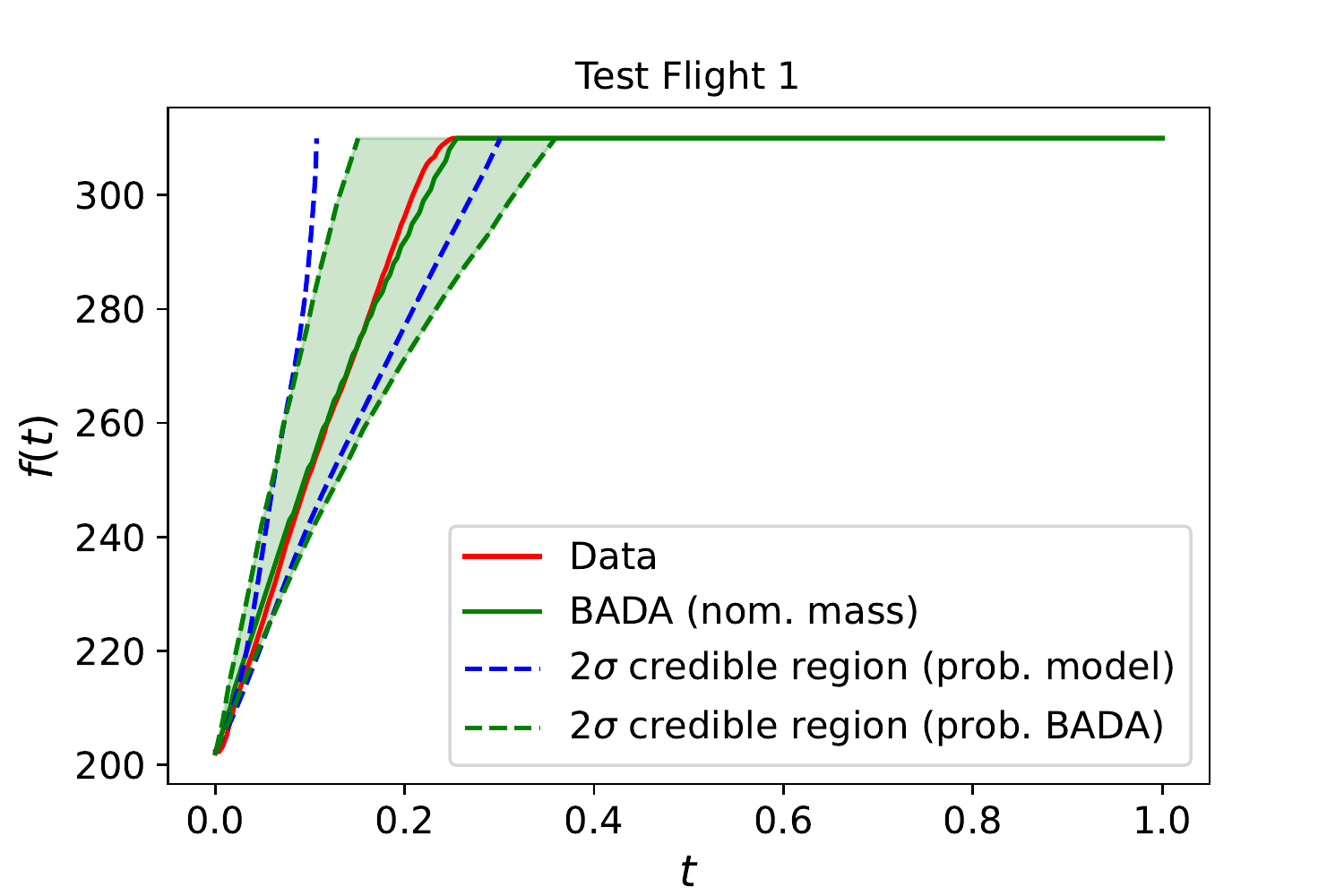}
\includegraphics[width=0.49\textwidth]{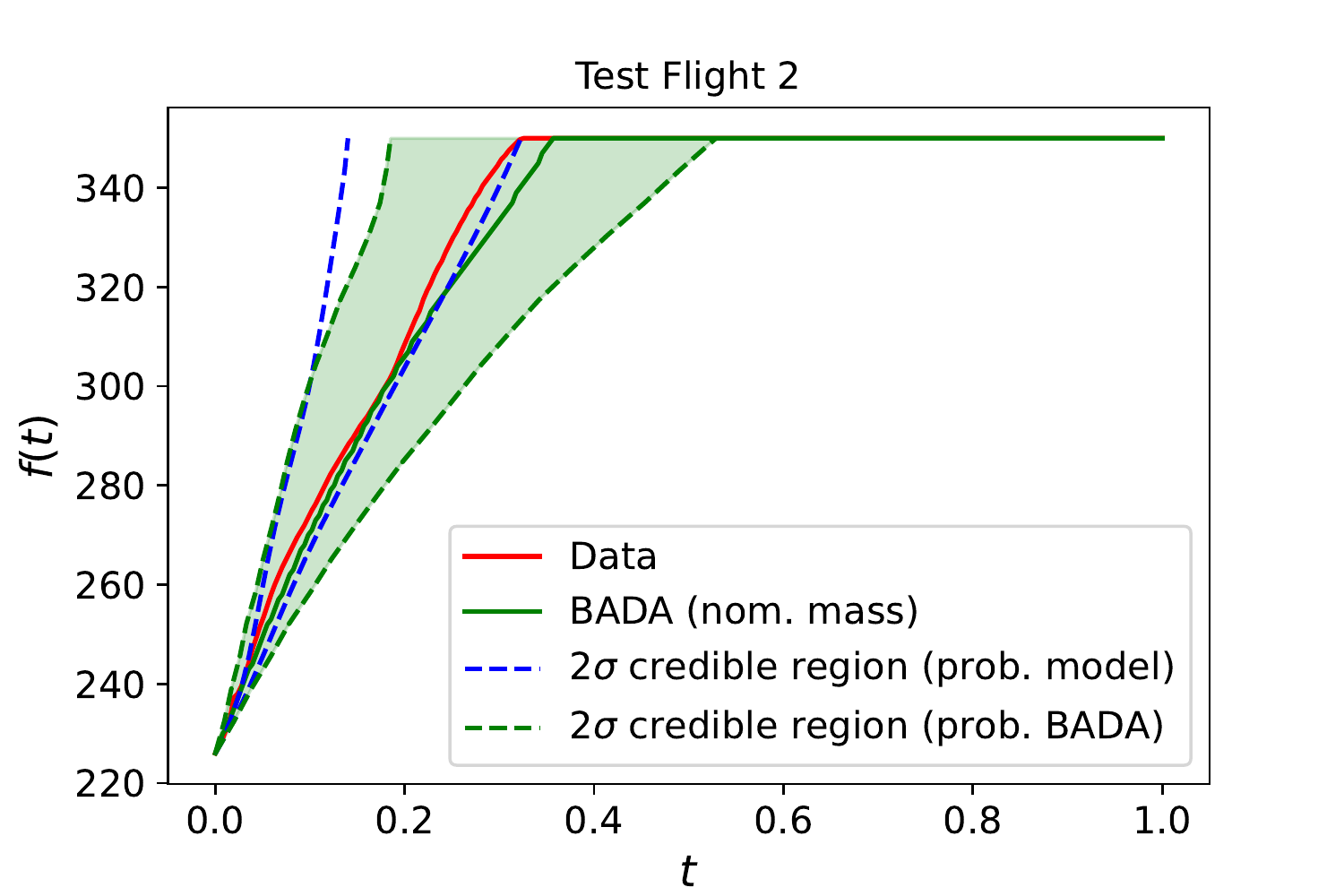}
\includegraphics[width=0.49\textwidth]{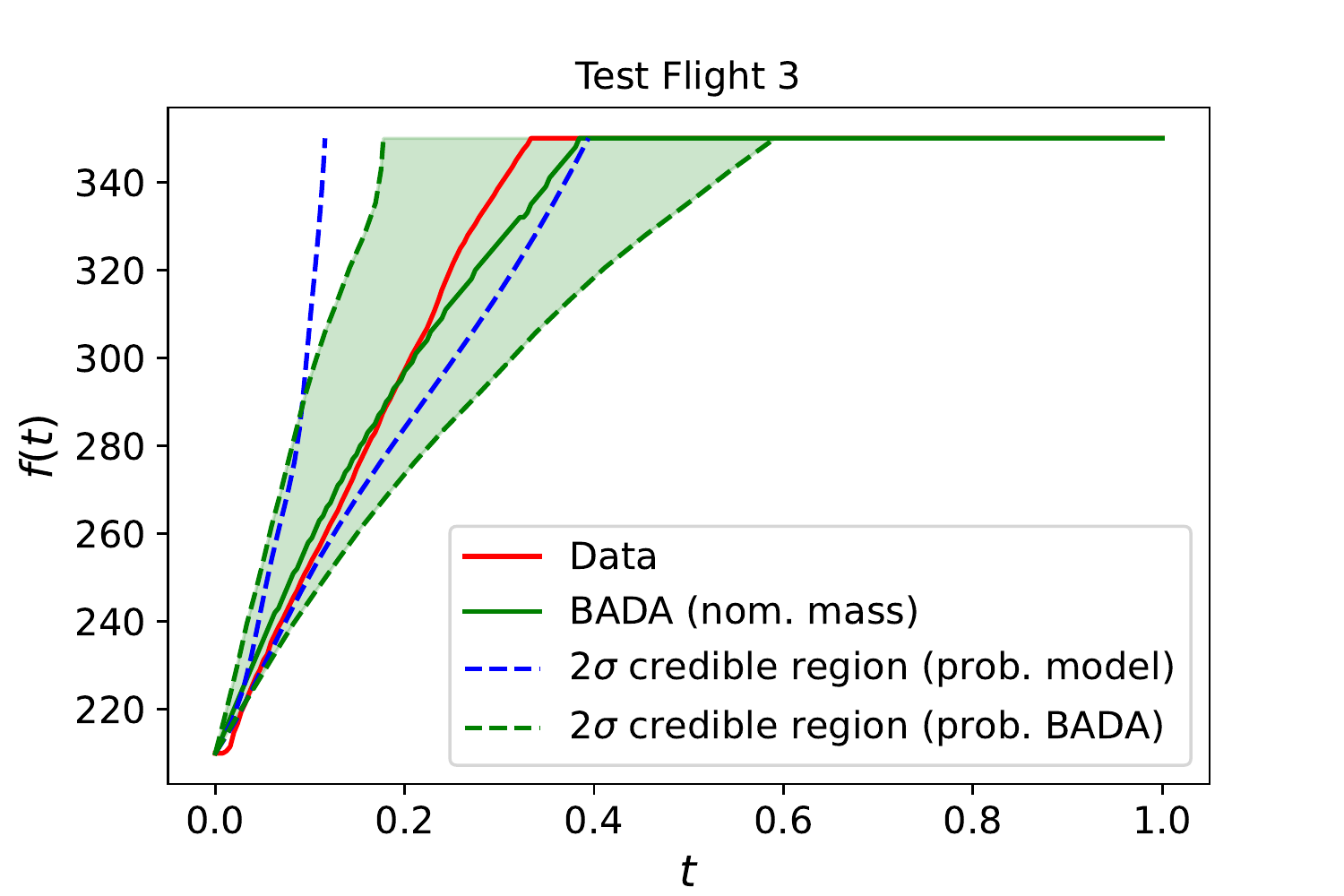}
\includegraphics[width=0.49\textwidth]{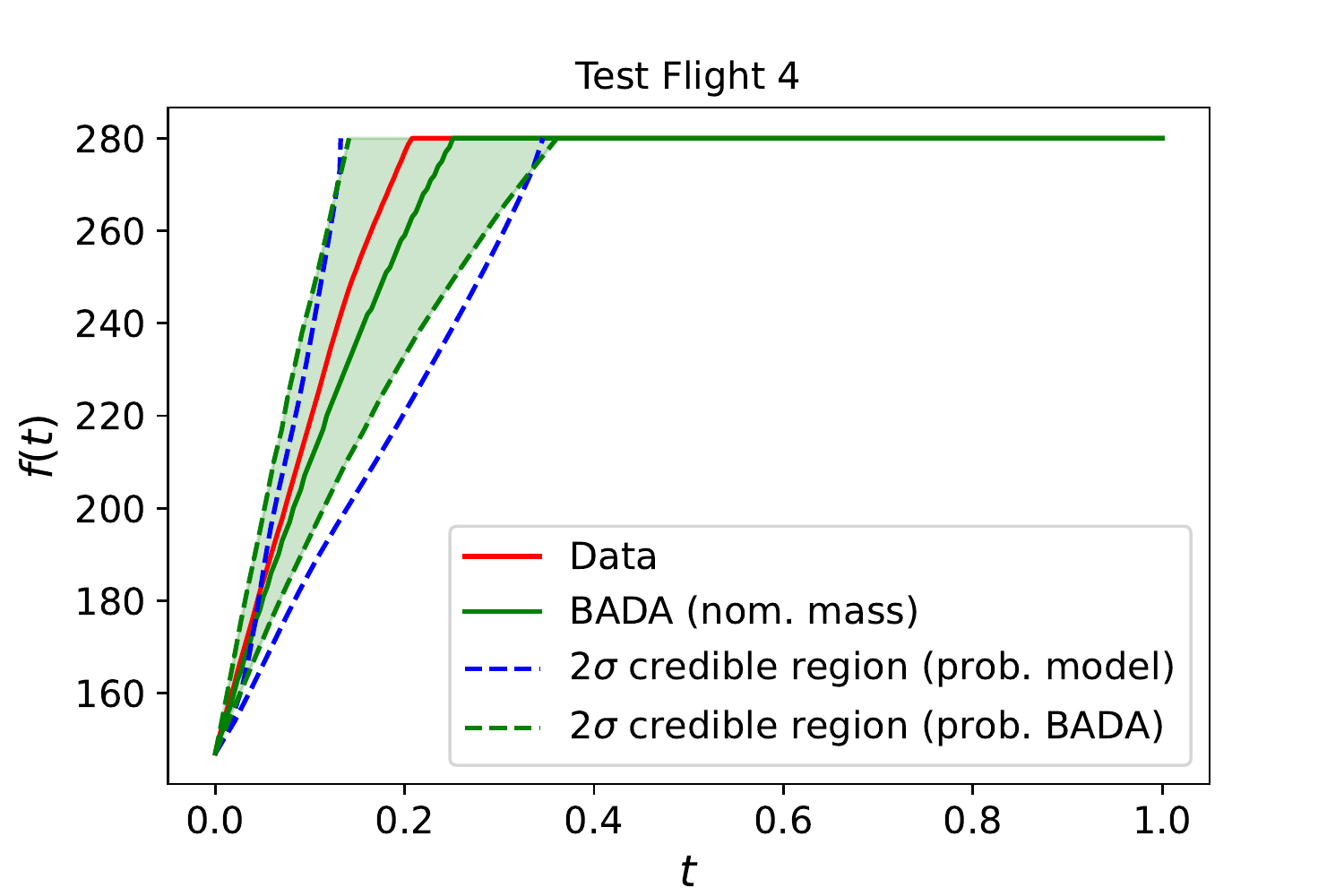}
\includegraphics[width=0.49\textwidth]{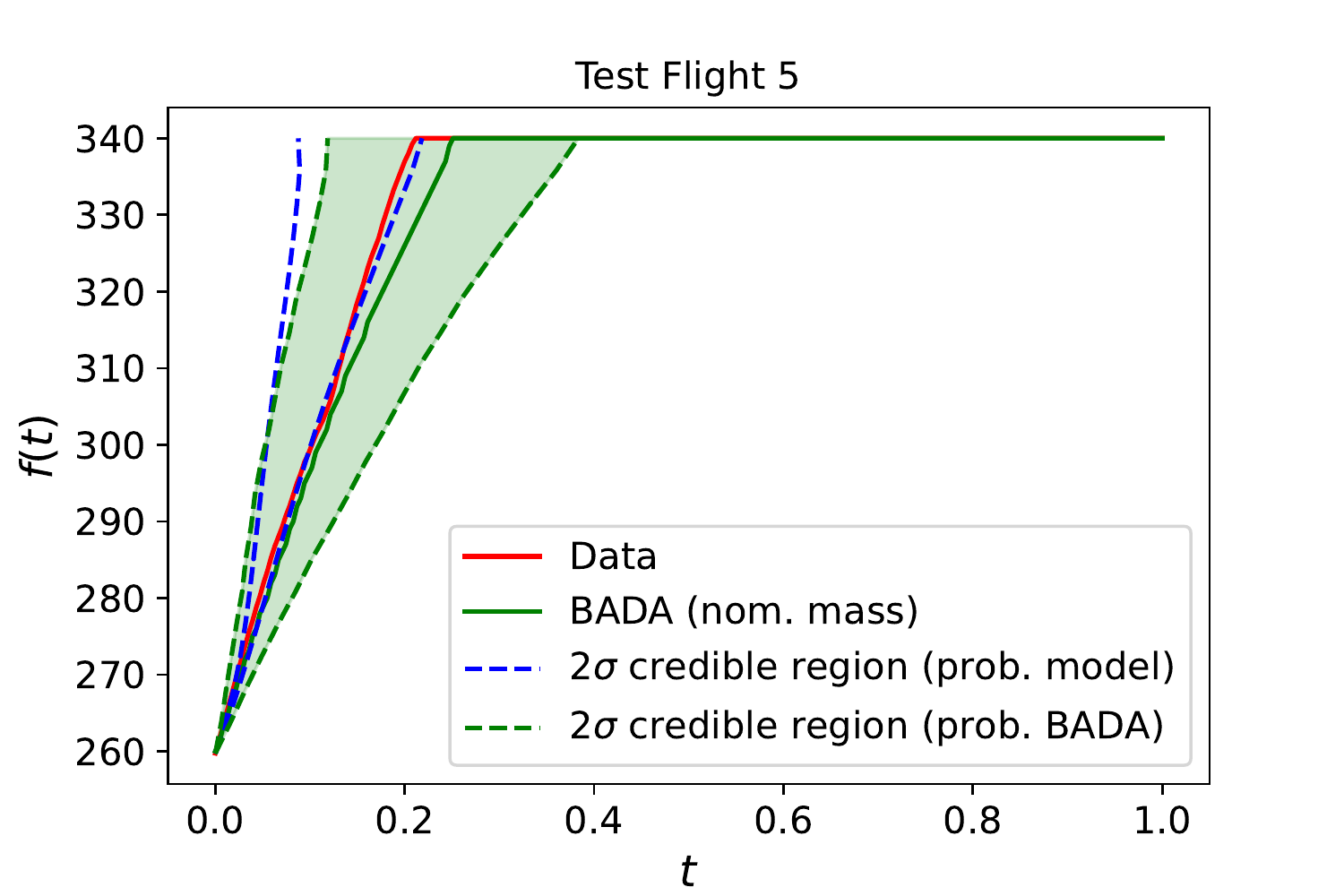}
\includegraphics[width=0.49\textwidth]{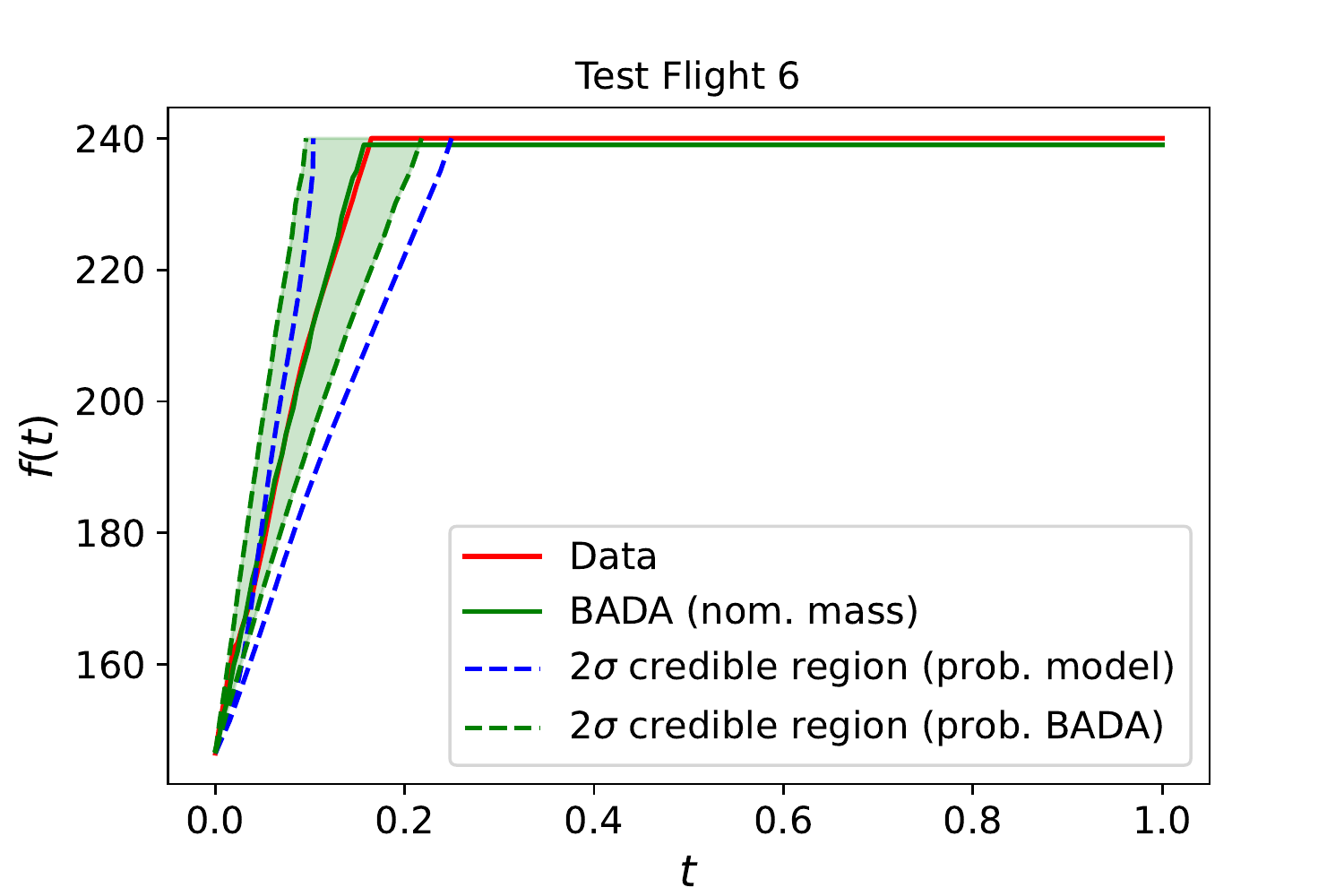}
\includegraphics[width=0.49\textwidth]{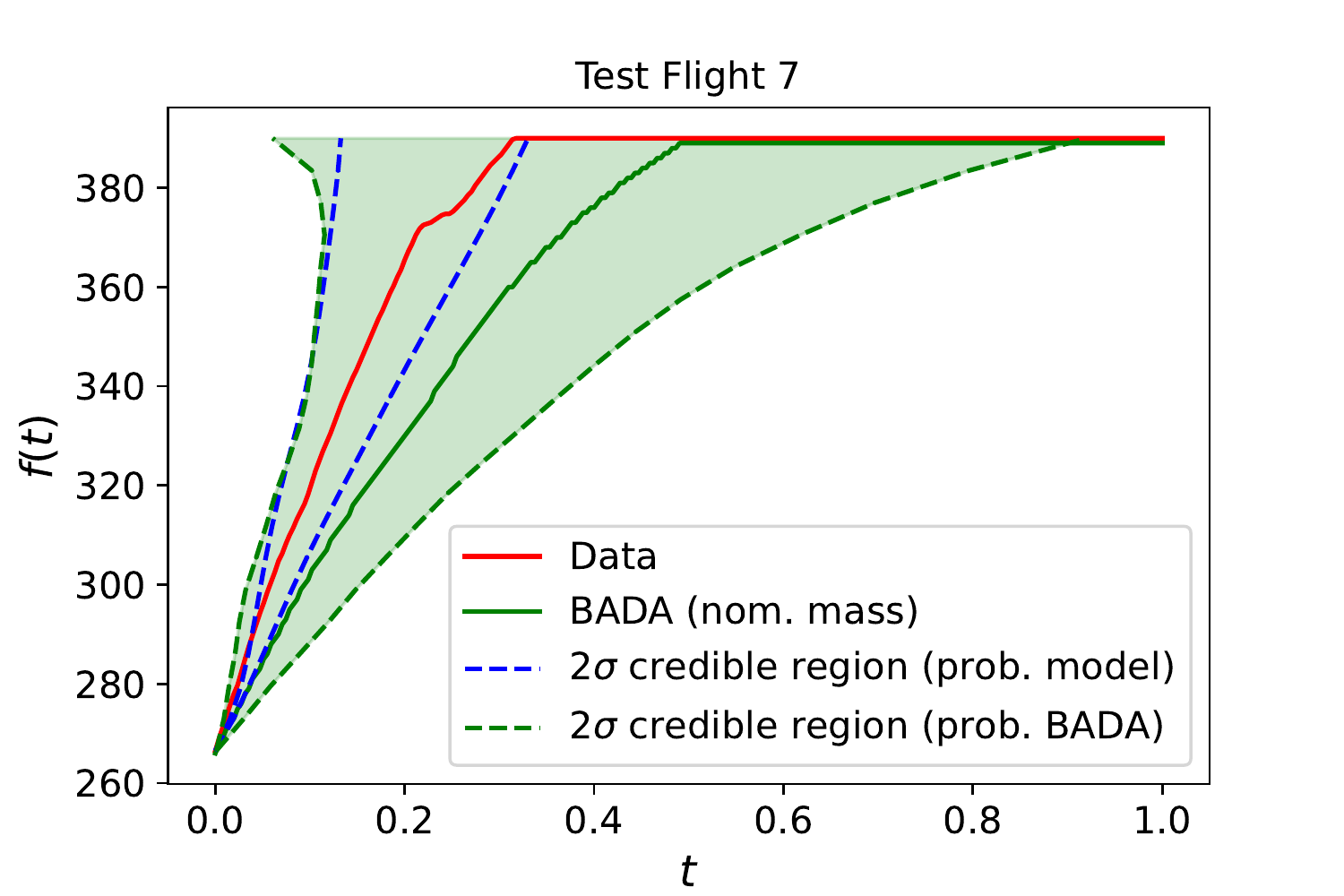}
\includegraphics[width=0.49\textwidth]{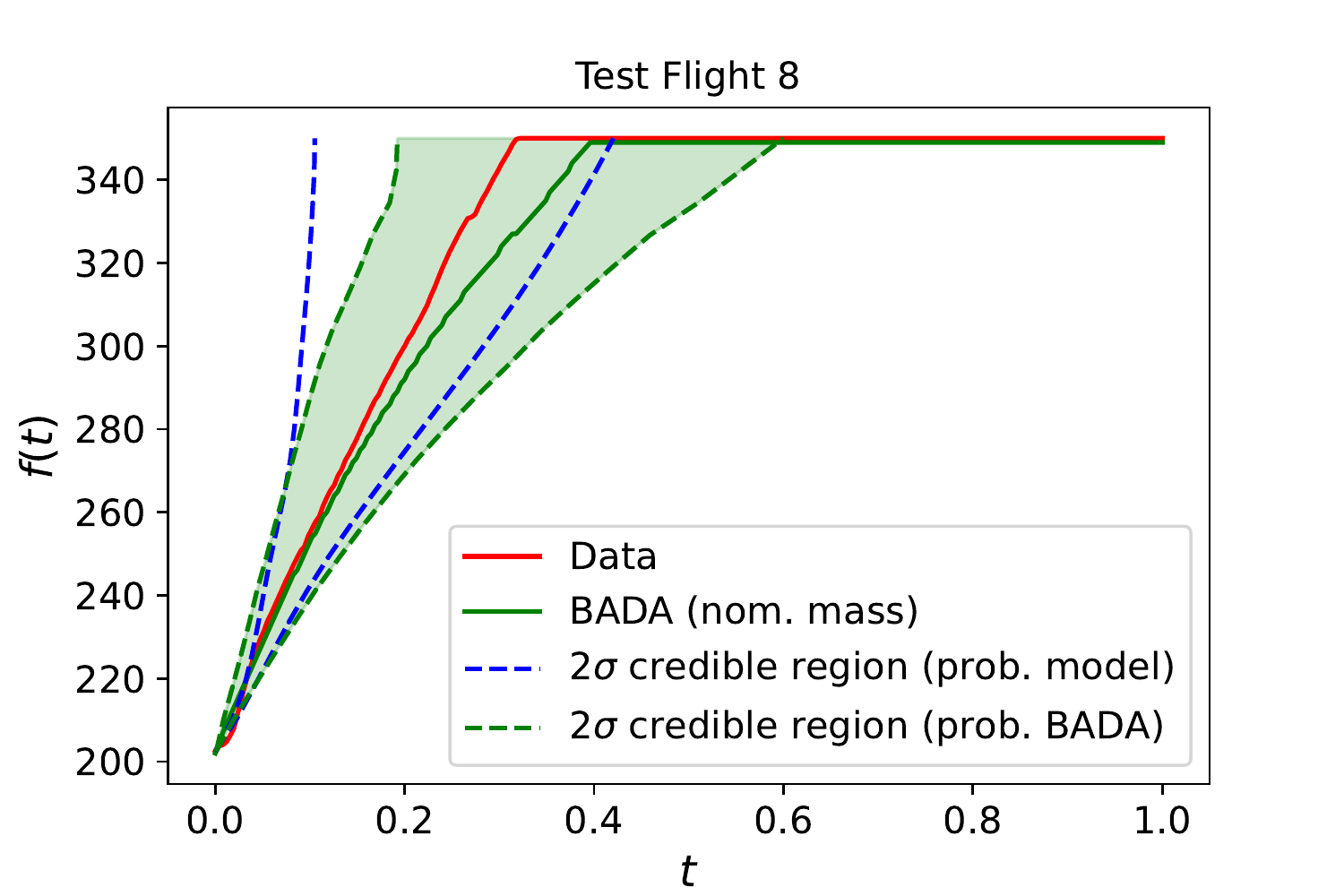}

\end{center}
\caption{Visual of the predicted trajectories for the same eight test points using the probabilistic implementation of BADA. The observed trajectory (red), is compared against the BADA run using nominal mass (green). The credible interval of the model is also plotted (dashed green) versus that of the proposed model (dashed blue).} 
\label{fig:test_points_probBADA}
\end{figure}

\begin{figure}
\begin{center}
\includegraphics[width=0.49\textwidth]{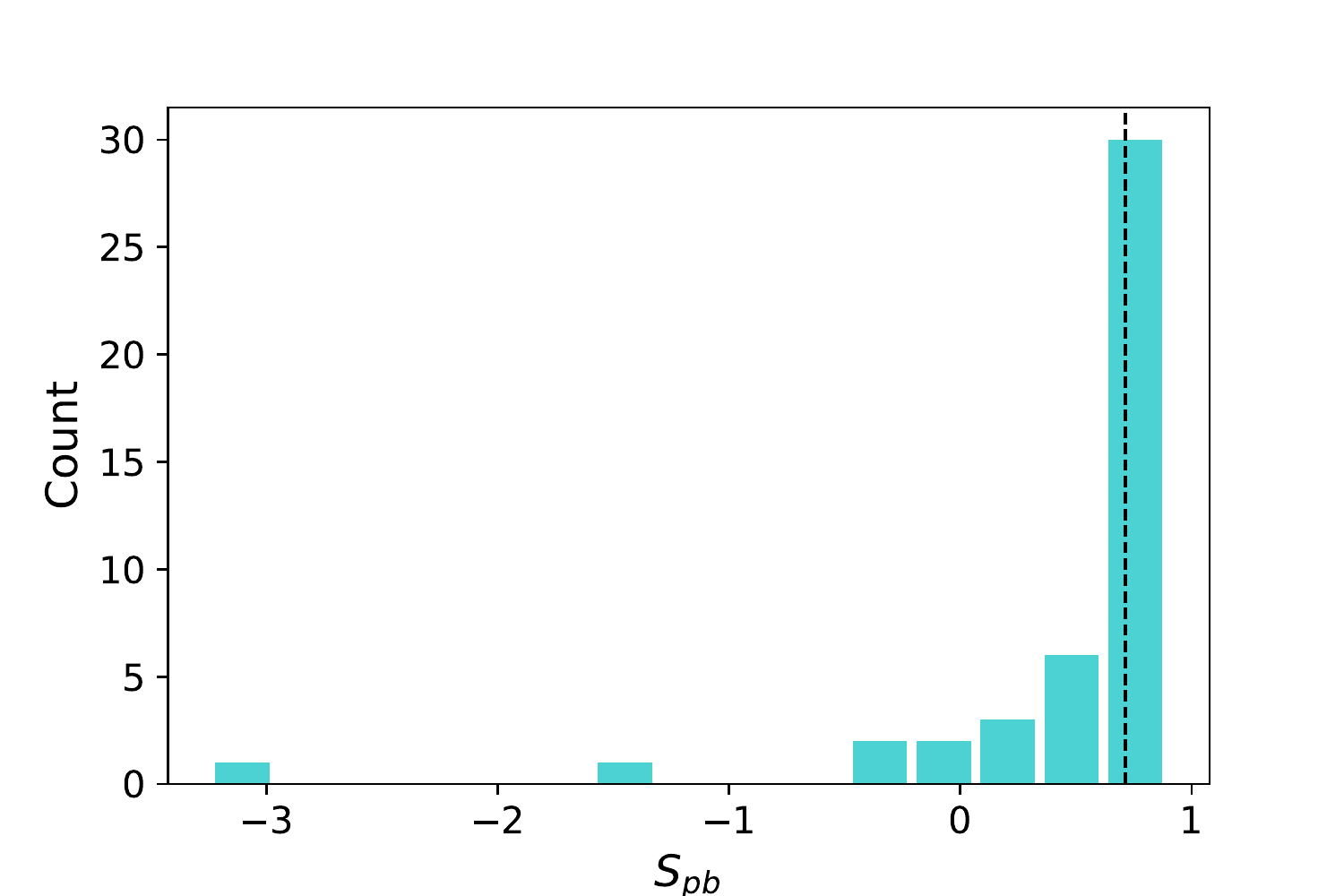}
\includegraphics[width=0.49\textwidth]{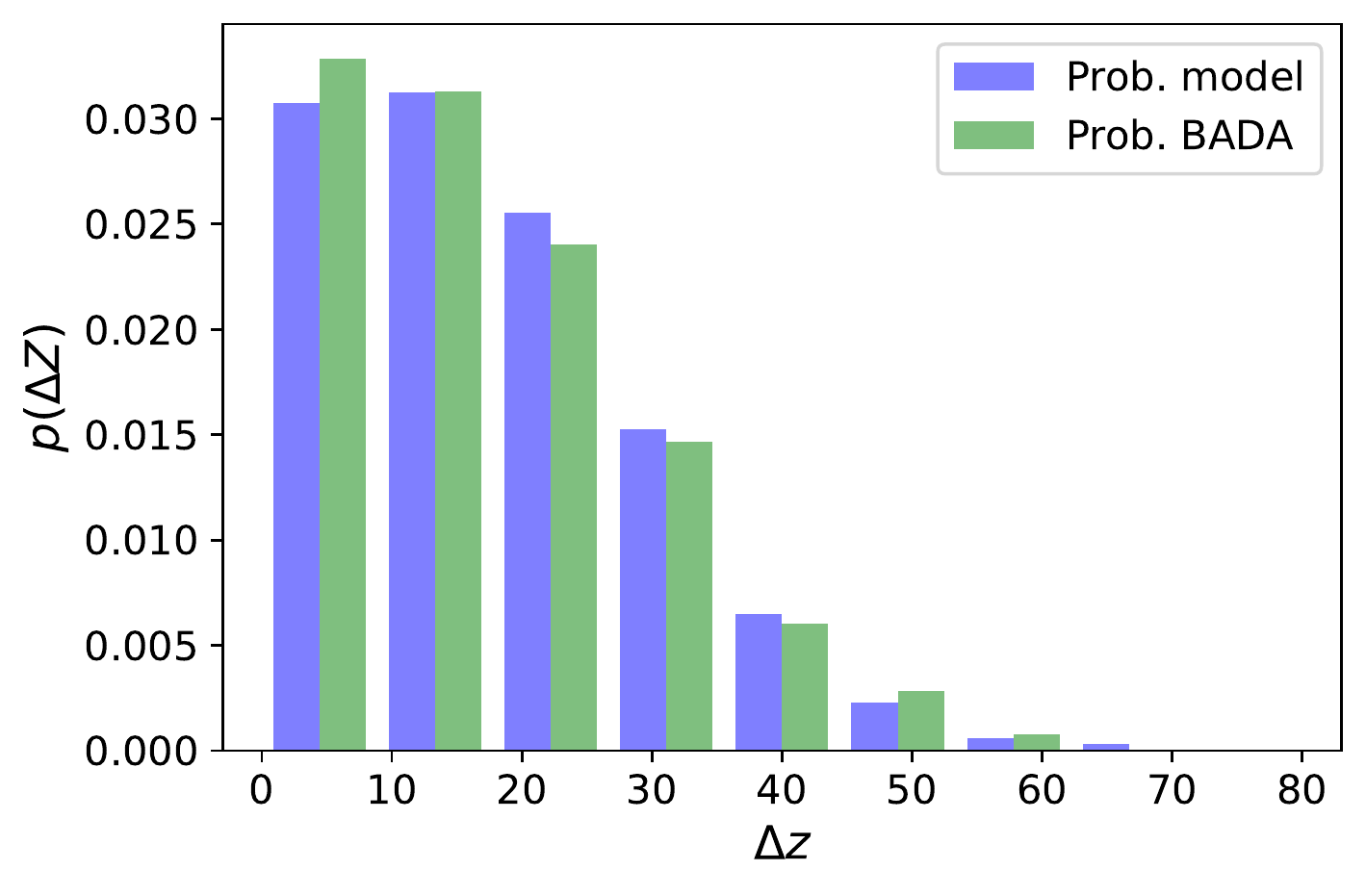}
\end{center}
\caption{Left: Histogram displaying the distribution of skill scores for the probabilistic model, baselined by a probabilistic implementation of BADA. Right: Normalised histograms for the discrepancies in the flight level at the end of climb for the samples.}
\label{fig:skill_hist_pBADA}
\end{figure}

Having compared the probabilistic model against BADA, run using the nominal mass value for the aircraft, in this section we make a comparison between the proposed model and a simple probabilistic implementation of BADA. Given the sensitivity of the BADA predictions to the mass of the aircraft, we formulate a simple probabilistic model for BADA in which the mass of the aircraft is sampled from a uniform distribution $U(m_{min}, m_{max})$, where $m_{min}$ and $m_{max}$ are the minimum and maximum masses associated with that aircraft in the BADA model. As before, we evaluate the two probabilistic models using the arrival time and offset in flight levels at the arrival time as performance indicators. In this case, the relative skilfulness of the proposed probabilistic model, baselined by the probabilistic BADA prediction, is denoted:
\begin{align}
    S_{pb}^{(i)}=1-\frac{CRPS^{(i)}}{CRPS^{(i)}_{pb}},
\end{align}
where $S_{pb}$ refers to the relative skillfulness, baselined by the probabilistic implementation of BADA, with associated CRPS score $CRPS_{pb}$. The left panel of Figure \ref{fig:skill_hist_pBADA} indicates the relative skilfulness for each trajectory in the test set. The average skill score of +68.25\% was higher than when the probabilistic model was baselined against the deterministic model evaluation. This is likely due to the much higher variance in the sampled trajectories from the probabilistic BADA, as can be seen in Figure \ref{fig:test_points_probBADA}. The CRPS score penalises overly-conservative forecasts. In Figure~\ref{fig:test_points_probBADA} the $2\sigma$ credible region from the probabilistic model is compared to the credible region from the probabilistic BADA model, with the data-driven bound from the probabilistic model being noticeably tighter. These confidence bounds were assessed using the Root Mean Squared Error of Calibration (RMSEC) \cite{ML_base, tran2020methods, kuleshov} and the sharpness. The RMSEC quantifies how well calibrated the two probabilistic TP models are, while the sharpness quantifies the concentration of the uncertainty. Ideally, we wish for a data-driven model to provide a sharper forecast than BADA, subject to calibration. Here we find that the data-driven confidence intervals are 35.52\% sharper, although the calibration error is increased by 4.58\%. To determine these values, a set of 20 intermediate flight levels were chosen for each test flight and the mean and variance of the arrival time were computed.

Finally, the right panel of Figure \ref{fig:skill_hist_pBADA} displays the normalised histograms for $\Delta z$. Note that in this case $\Delta z$ was calculated for each of the posterior samples, rather than the mean estimate, hence the chart for the probabilistic model differs from that in Figure \ref{fig:skill_hist}. The data-driven posterior samples were found to have a much lower average discrepancy.


\begin{table}[!htbp]
\centering
\caption{The performance of the probabilistic model, bench-marked against BADA for various metrics, including: the Mean Absolute Error (MAE), relative skill score for predicting the end of climb, flight level discrepancy at the end of climb ($\Delta z$), and Root Mean Squared Error of Calibration (RMSEC). All quantities are averaged over the 45 climbs in the test dataset. Arrows indicate whether higher or lower values of that metric are preferable.}
\begin{tabular}{*6c}
\toprule
Method & MAE $\downarrow$ & $S$ (\%) $\uparrow$  & {$\Delta z$ $\downarrow$} & RMSEC $\downarrow$ & Sharpness $\downarrow$\\
\midrule
Prob. Model & 1.840 & - & 13.70 (mean) & 0.1875 & 0.02433 \\
& & & 17.71 (post. samp.) & & \\
BADA (nom.) & 2.322 & 32.86 & 11.96 & - & -\\
BADA (prob.) & 1.927 & 68.25 & 17.24 & 0.1923 & 0.0366 \\
\bottomrule
\end{tabular}
\label{Table:1}
\end{table}

\begin{table}[!htbp]
\centering
\caption{Skill scores for the eight plotted test flights and the relative sharpness of the methods}
\begin{tabular}{*4c}
\toprule
Test flight & \multicolumn{2}{c}{$S$ $\uparrow$} & Relative sharpness $\downarrow$\\
& BADA (nom.) & BADA (prob.) & \\
\midrule

1 & -6.529 & 0.7059 & 0.8895 \\
2 & -0.9591 & 0.5809 & 0.5576\\
3 & 0.0421 & 0.7427 & 0.7251\\
4 & 0.4718 & 0.7657 & 1.087\\
5 & 0.04151 & 0.7095& 0.4622\\
6 & -0.3924 & 0.8674& 1.137\\
7 & 0.6400 & 0.4524 & 0.2869\\
8 & 0.6230 & 0.8154 & 0.8494\\
\bottomrule
\end{tabular}
\label{Table:2}
\end{table}


\newpage
\section{Conclusion}
Accurate trajectory prediction for an aircraft in climb is difficult due to the presence of significant epistemic uncertainty. We have proposed a probabilistic model that employs Gaussian Process Emulators to give a fast approximation to a trajectory that is functional and monotonic. A functional approach allows tempo-spatial correlations in the training data to be exploited. In this paper the proposed method has been baselined against a deterministic TP model, widely used in industry, in addition to a probabilistic version that exploits the available knowledge of the aircraft's mass. Using a test dataset of unseen data, it was found that the proposed model outperformed both the deterministic and probabilistic versions of BADA, with performance quantified by several metrics including the relative skill of the model in predicting the arrival time of an aircraft at its target flight level.

In this paper we have restricted the application of the model to TP for climbing aircraft, although we note that the method is general and could be applied to any class of problem where a fast approximation for monotonic functional outputs is required. Finally, in its presented form the method is completely data-driven. However, we note that a deterministic model could be used to inform the model by acting as a mean function for the GPEs. In such an instance, it would be necessary to find a mapping such that solutions of the deterministic model could be expressed in the coefficient space of the Ramsay framework.
\enlargethispage{20pt}






\bibliographystyle{unsrtnat}

\appendix

\section{Parameter estimation using SGD}
Trajectory data in the training set is collected from radar observations of an aircraft and is therefore discretised according to the number of radar `blips' during the climb or descent. $D$ is more realistically expressed as $D=\{\boldsymbol{x}^{(i)}, f^{(i)}(t_j),\, j=1\dots n_b^{(i)}\}$ where $i=1\dots n_d$ (there is no guarantee that each sampled trajectory will have the same number of radar observations). For the $i$\textsuperscript{th} trajectory we wish to estimate the parameters in the monotonic representation in \eqref{mono1} and \eqref{mono2}. This can be cast as the optimisation problem:
\begin{align}
    \underset{\beta_1, a_0, {a}, {b}}{\rm{min}}\;\mathcal{L}_i (\beta_1, a_0, \boldsymbol{a}, \boldsymbol{b}),
\end{align}
where we define the loss function for the $i$\textsuperscript{th} trajectory, $\mathcal{L}_i$, as the residual sum of squares (RSS) loss between the observations and the functional representation, i.e.:
\begin{align}
    \mathcal{L}_i (\beta_1, a_0, \boldsymbol{a}, \boldsymbol{b})=\sum_{j=1}^{n_b^{(i)}}\{f^{(i)}_j-f(t_j|\beta_1, a_0, \boldsymbol{a}, \boldsymbol{b})\}^2.
\end{align}
Analytic expressions for the derivatives of the loss function are available:
\begin{align}
    \frac{\partial \mathcal{L}_i}{\partial f_j}=-2\{f_j^{(i)}-f_j\},
\end{align}
where we have defined $f_j=f(t_j|\beta_1, a_0, \boldsymbol{a}, \boldsymbol{b})$ for ease of notation. Using the chain rule we can find derivatives for each of the parameters:
\begin{align}
    & \frac{\partial \mathcal{L}_i}{\partial \beta_1}=\sum_{j=1}^{n_b^{(i)}}\frac{\partial \mathcal{L}_i}{\partial f_j}\frac{\partial f_j}{\partial \beta_1}=-2\sum_{j=1}^{n_b^{(i)}}\{f_j^{(i)}-f_j\}\int_{\tau_0}^{t_j} \text{exp}\int_{\tau_0}^s w(u) \text{d}u\, \text{d}s\\
 & \frac{\partial \mathcal{L}_i}{\partial a_0}=\sum_{j=1}^{n_b^{(i)}}\frac{\partial \mathcal{L}_i}{\partial f_j}\frac{\partial f_j}{\partial a_0}=-2\beta_1\sum_{j=1}^{n_b^{(i)}}\{f_j^{(i)}-f_j\} \{\text{exp}(t_j-\tau_0)-1\} \\
 & \frac{\partial \mathcal{L}_i}{\partial \boldsymbol{a}_k}=\sum_{j=1}^{n_b^{(i)}}\frac{\partial \mathcal{L}_i}{\partial f_j}\frac{\partial f_j}{\partial \boldsymbol{a}_k}=-2\beta_1\sum_{j=1}^{n_b^{(i)}}\{f_j^{(i)}-f_j\} \int_{\tau_0}^{t_j} \text{exp}\int_{\tau_0}^s \text{cos} (2\pi k u) \text{d}u\, \text{d}s\\
  & \frac{\partial \mathcal{L}_i}{\partial \boldsymbol{b}_k}=\sum_{j=1}^{n_b^{(i)}}\frac{\partial \mathcal{L}_i}{\partial f_j}\frac{\partial f_j}{\partial \boldsymbol{b}_k}=-2\beta_1\sum_{j=1}^{n_b^{(i)}}\{f_j^{(i)}-f_j\} \int_{\tau_0}^{t_j} \text{exp}\int_{\tau_0}^s \text{sin} (2\pi k u) \text{d}u\, \text{d}s
\end{align}

Adagrad is a gradient based optimisation algorithm used to update the parameters with a variable learning rate \cite{adagrad}. At the $m$\textsuperscript{th} iteration the
set of parameters, $\boldsymbol{y}=\{\beta_1, a_0,\boldsymbol{a}, \boldsymbol{b}\}$ is adjusted according to:
\begin{align}
    \boldsymbol{y}_{l,m}=\boldsymbol{y}_{l,m-1}-\frac{\eta}{\sqrt{G_{m_{ll}}+\rho}}\frac{\partial \mathcal{L}_i}{\partial \boldsymbol{y}_{l,m-1}},
\end{align}
where $G_m$ is a diagonal matrix, in which the element $G_{m_{ll}}$ contains the sums of the squares of the past gradients for the $l$\textsuperscript{th} parameter. The constant $\rho$ is a smoothing constant used to ensure numerical stability when the gradients are small, while the learning rate $\eta\sim\mathcal{O}(10^{-2})$ is set by the user. Here we used $\eta=0.02$.

SGD is run to convergence for each trajectory in $D$, defined by a stopping criterion for the relative improvement of $\mathcal{L}$, yielding the set of parameters $D_y=\{\hat{\boldsymbol{y}}^{(i)}\},\,i=1,\dots,n_d$. In what has been presented the search for the parameters is unconstrained, although we note that performance constraints of the aircraft, if known, could be expressed through a bound on $w$. Such constraints could be expressed through a feasibility function $g(w)$, yielding the modified optimisation problem:
%
%
%
%
%
%
\begin{align}
    &\underset{\beta_1, a_0, {a}, {b}}{\rm{min}}\;\mathcal{L}_i (\beta_1, a_0, \boldsymbol{a}, \boldsymbol{b}) \\
    &\textrm{such that}\; g(w|\beta_1, a_0, \boldsymbol{a}, \boldsymbol{b}, \boldsymbol{x}_{per})>0, \nonumber
    \label{eq:phys_constraint}
\end{align}
where $\boldsymbol{x}_{per}$ collects the additional performance related data required to evaluate $g(\cdot)$.



\section{Sub-algorithm to truncate Principal Component representation}
Section 2\ref{section:mono} discusses how PCA is used to rotate $D_y$ into a basis with statistically independent components. A further advantage of using the PCA representation is that the PCs are ranked in terms of the explained variance, allowing PCs that do not contribute significantly to be neglected. Typically, this is done through an analysis of the explained variance ratio (see, e.g. \cite{cum_exp_ratio}). However, in this case it is difficult to quantify how variance in $\boldsymbol{y}$ scales with variance in the generated trajectories when passed through the monotonic framework. For this reason, we propose a simple adaptive method for selecting $n_c$, based on the error associated with reconstructing the trajectories in $D$ after they have been projected into the PC basis. We denote this quantity $\epsilon_r$, calculated through:
\begin{align}
    \epsilon_r=\sum_{i=1}^{n_d}\sum_{j=1}^{n_b^{(i)}} \bigg(f^{(i)}(t_j)-f(t_j|\boldsymbol{x}^{(i)}, \boldsymbol{y})\bigg)^2,
\end{align}
where
\begin{align}
    \boldsymbol{y}=\sum_{k=1}^{n_c}\boldsymbol{\alpha}_k\boldsymbol{\psi}_k.
\end{align}

\begin{figure}[H]
\begin{center}
\includegraphics[width=0.6\textwidth]{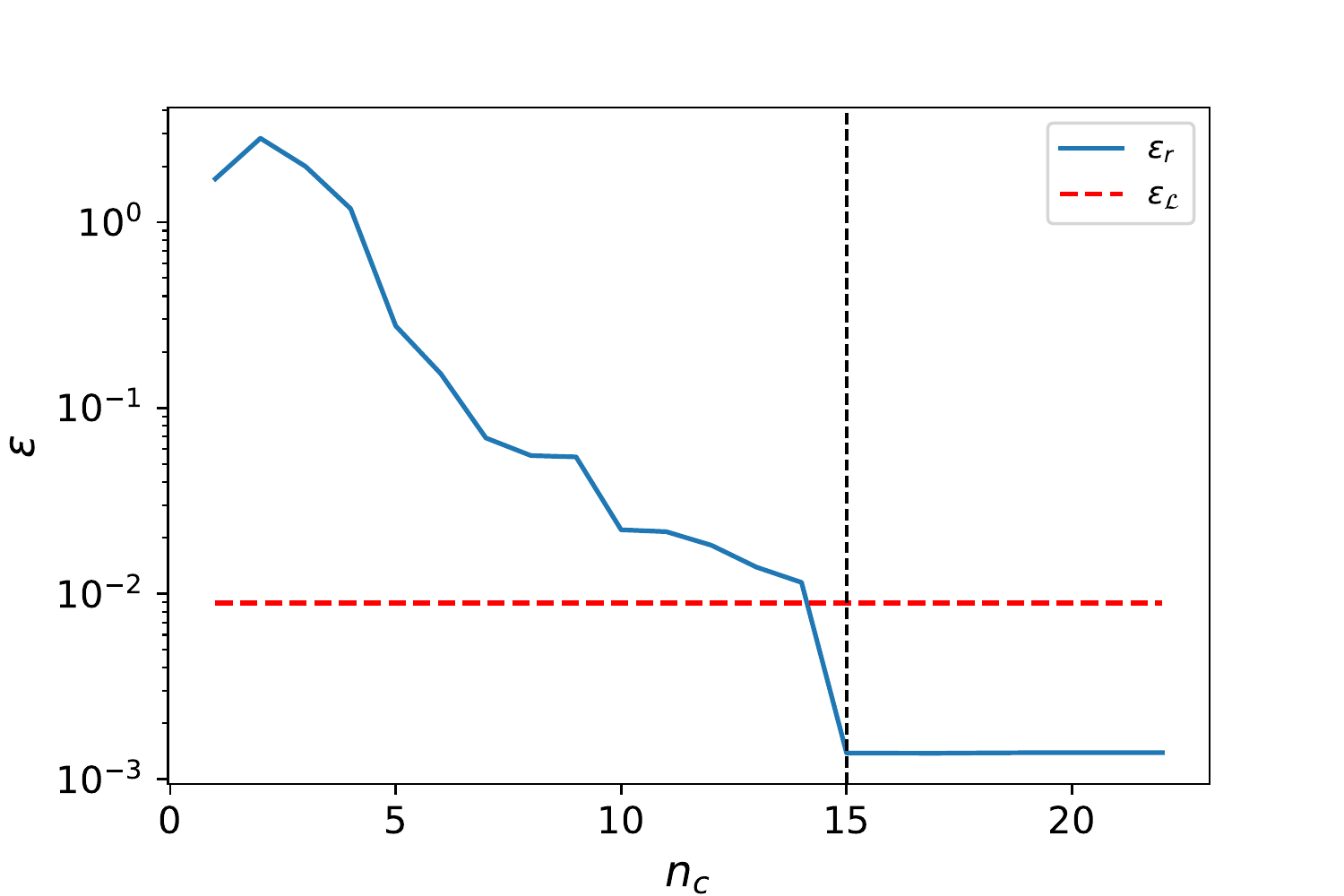}
\end{center}
\caption{Plot displaying variation in $\epsilon_r$ and $\epsilon_\mathcal{L}$ (scaled by $n_d$ and number of radar blips) with number of principal components. The vertical dashed line indicates the $n_c$ chosen by the adaptive process.}
\label{fig:adaptive}
\end{figure}

We remove PCs until a terminal condition on $\epsilon_r$ is met. Here we use the relative ratio of the reconstruction error to the error introduced in the parameter estimation, $\epsilon_\mathcal{L}$, defined as $\epsilon_\mathcal{L}=\sum_{i=1}^{n_d} \mathcal{L}_i$. In the implementation of the algorithm described in Section 3, we define this stopping criterion as $\frac{\epsilon_r}{\epsilon_\mathcal{L}}\geq 1.01$. This is illustrated in Figure \ref{fig:adaptive}.

\end{document}